\documentclass[prl,aps,amsfonts,twocolumn,superscriptaddress,notitlepage,floatfix]{revtex4-1}
\usepackage{amsmath,amssymb}
\usepackage{graphicx}
\usepackage{epsfig, color}
\usepackage{wasysym}
\usepackage{amsfonts}
\usepackage{bm}
\usepackage{enumerate}
\usepackage{bbold}
\usepackage{xcolor}
\usepackage[normalem]{ulem}
\usepackage{cancel}
\usepackage{multibib}
\usepackage{comment}

\usepackage{mathtools}

\usepackage{latexsym}
\usepackage[breaklinks,colorlinks = true,linkcolor = blue,urlcolor=cyan,citecolor=red,linktocpage=true]{hyperref}

\newcommand{\be}{\begin{eqnarray}}
\newcommand{\ee}{\end{eqnarray}}
\newcommand{\bent}{\begin{equation*}}
\newcommand{\eent}{\end{equation*}}

\usepackage{pdfpages} 
\makeatletter
\AtBeginDocument{\let\LS@rot\@undefined}
\makeatother

\newcommand{\vpd}[0]{\vphantom{\dagger}}

\def\ket#1{{|#1\rangle}}
\def\bra#1{{\langle #1 |}}
\newcommand{\ident}[0]{ \mathbb{1}}

\newcommand{\expval}[1]{ \left\langle #1 \right\rangle}

\newcommand{\tr}[1]{{\rm tr} \left[ #1 \right]}
\newcommand{\trace}{{\rm tr}}

\def\gate{\mathcal{U}}
\def\Haar{U}
\def\Floq{\mathcal{F}}

\def\Ham{H}
\def\Tmat{\mathcal{T}}
\def\Tgate{T}
\def\observ{\mathcal{O}}

\def\tth{t^{\vpd}_{\rm Th}}
\def\theis{t^{\vpd}_{\rm Heis}}

\def\Hdim{\mathcal{D}}
\def\Ldim{q}
\def\ancilla{d}

\def\gap{\Delta}
\def\charge{\mathfrak{q}}
\def\totcharge{Q}
\def\Corr{\mathcal{C}}
\def\numsec{\mathcal{N}}

\newcommand\particle{{\color{black}\bullet}\mathllap{\circ}}
\newcommand\hole{\circ}

\newcommand{\opket}[1]{\left| #1 \right)}

\newcommand{\opinprod}[2]{\left( #1 \middle| #2 \right)}

\newcommand{\opmatel}[3]{\left( #1 \middle| #2 \middle| #3 \right)}

\begin{document}

\title{Subdiffusion and many-body quantum chaos with kinetic constraints}

\author{Hansveer Singh}
\affiliation{Department of Physics, University of Massachusetts, Amherst, Massachusetts 01003, USA}
\author{Brayden A. Ware}
\affiliation{Department of Physics, University of Massachusetts, Amherst, Massachusetts 01003, USA}
\author{Romain Vasseur}
\affiliation{Department of Physics, University of Massachusetts, Amherst, Massachusetts 01003, USA}
\author{Aaron J. Friedman}
\affiliation{Department of Physics and Center for Theory of Quantum Matter, University of Colorado, Boulder, Colorado 80309, USA}
\date{\today}

\begin{abstract}
We investigate the spectral and transport properties of many-body quantum systems with conserved charges and kinetic constraints. Using random unitary circuits, we compute ensemble-averaged spectral form factors and linear-response correlation functions, and find that their characteristic time scales are given by the inverse gap of an effective Hamiltonian---or equivalently, a transfer matrix describing a classical Markov process. Our approach allows us to connect directly the Thouless time, $t^{\vphantom{\dagger}}_{\rm Th}$, determined by the spectral form factor, to transport properties and linear response correlators. 
Using tensor network methods, we determine the dynamical exponent, $z$, for a number of constrained, conserving models. We find universality classes with diffusive, subdiffusive, quasilocalized, and localized dynamics, depending on the severity of the constraints. In particular, we show that quantum systems with ``Fredkin'' constraints exhibit anomalous  transport with dynamical exponent $z \simeq 8/3$.

\end{abstract}

\maketitle

\noindent \emph{Introduction.---}\,
Recent years have seen substantial progress in understanding how isolated quantum systems thermalize under their own dynamics.~
The eigenstate thermalization hypothesis (ETH) \cite{ETH1,ETH2} proposes that entanglement between subsystems allows for local equilibration: Generic unitary evolution scrambles local quantum information into highly nonlocal degrees of freedom, which are inaccessible to local observables. Early tests of ETH~\cite{PhysRevLett.98.050405,Rigol2008kq,santos2010onset,borgonovi2016quantum,Rigol} relied on 
small scale numerics and extensions of integrable models, which are fine tuned; understanding the universal aspects of quantum chaotic dynamics requires a more general approach.

A hallmark of chaotic systems is that they dynamically forget as much information about their past as symmetries allow. Hence, the salient features of chaotic systems are well captured by replacing the microscopic model with a random matrix with the same symmetries. Random unitary circuits (RUCs) invoke the potency of random matrix theory (RMT) while also introducing spatial locality, with the system evolved by a brickwork ``circuit'' of $\ell$-site gates \cite{YoshidaCBD,ProsenRMTChaosPRX,NahumRUC1,NahumOperator,RUCNCTibor,CiracRUFC,CDLC1,Kos_2021,U1FRUC}. RUCs are fully generic, and their study has elucidated the universal dynamics of chaotic quantum systems: Entanglement grows linearly until saturating to a volume law, with fluctuations in the KPZ universality class \cite{NahumRUC1}; operator fronts (and out-of-time-ordered correlation functions) propagate ballistically and broaden diffusively \cite{NahumOperator,RUCNCTibor,CDLC1}, etc. 

However, these RUCs are designed to be featureless; an interesting question is how these properties change as one reintroduces other physical ingredients, such as symmetries. With conserved charges, one can consider transport; for typical $U(1)$ symmetry, one expects conserved charges to diffuse \cite{RUCconTibor,RUCconVedika,ShenkerRMT,U1FRUC}. Operators that overlap with conserved quantities are expected to have slower dynamics, dominated by hydrodynamic modes. It is also interesting to study dynamics 
with more complicated symmetries or constraints~\cite{Valado_2016,PhysRevB.94.235122, PhysRevLett.121.085701, GopalakrishnanAutomata_2018, Lan_2018,Everest_2016,QuantumEast2020,Scherg_2021,PhysRevLett.122.250602,PhysRevResearch.2.033020,Huang_2020}. Fractons, e.g., are excitations in systems with charge and dipole conservation, which are constrained to move in pairs only \cite{SagarFracton1,SagarFracton2}. This higher-order symmetry can also be viewed as a \emph{constraint}; recent studies of fractons in the contexts of RUCs and hydrodynamics have found evidence for subdiffusion, with dynamical exponent $z=2(m+1)$, where $m$ is the highest conserved moment of charge~\cite{FractonSubdiffusionDim,FractonHydro,PhysRevLett.125.245303,morningstarKineticallyConstrainedFreezing2020,FractonSubdiffusion,glorioso2021breakdown,SanjayAmosFracton}.

In this Letter we analyze the general consequences of kinetic constraints on charge-conserving many-body quantum dynamics in one dimension. Kinetic constraints restrict the local rearrangements of charges and have been intensely studied as models of classical systems with glassy dynamics \cite{Valado_2016,QuantumEast2020,Scherg_2021,Lan_2018,Everest_2016,garrahan2010kinetically,GarrahanLectures,RitortKCM,KCMRydberg}. Depending on the 
locally forbidden rearrangements, adding constraints may anomalously slow down or completely freeze the process of thermalization. Using variations of RUCs, we probe how imposing constraints on generic quantum systems leads to new universality classes with slow dynamics. Using Floquet random circuits, in the limit of large on-site Hilbert space, we relate the scaling of the many-body Thouless time---the time scale for a system to show RMT spectral rigidity---with system size~\cite{CDLC1,CDLC2,bertini2018exact,ShenkerRMT,ShenkerRMT,FyodorovMirlin,CiracRUFC,U1FRUC,SwingleSFFHydro,SwingleSFFSSB,Kos_2021} to the inverse gap of the transfer matrix of a stochastic classical model; or equivalently, of an effective Hamiltonian, which lies at a Rokhsar-Kivelson (RK) point \cite{RK,SanjayAmosFracton,U1FRUC}. We show that the same 
transfer matrix also controls the dynamics of linear response correlators, providing a general relation between the Thouless time and transport \cite{ShenkerRMT}. Depending on the severity of the model's constraints, we find 
diffusive, subdiffusive, quasilocalized, and localized dynamics, and identify a new universality class of constrained $z \simeq 8/3$ ``Fredkin'' systems \cite{FredkinEntanglement,Korepin2016fredkin,Korepin2017fredkin,ChenFredkin2017,ChenFredkin2017a,Zhang2017fredkin,Udagawa2017fredkin,FredkinFragmentScar,ChenFredkin2020}
.

\noindent \emph{Spectral rigidity and transport correlators.---}\, A useful indicator of quantum chaos is \emph{level repulsion} \cite{ETH1,ETH2,ShenkerRMT}, characterized by an RMT distribution of the eigenvalues of the evolution operator~\cite{ShenkerRMT,ShenkerRMT,ProsenRMTChaosPRX,SwingleSFFHydro,CDLC1,CDLC2,U1FRUC,Kos_2021}. Periodically driven (Floquet) RUCs~\cite{CDLC1,CDLC2,CiracRUFC,U1FRUC}, afford such a spectrum, as time evolution follows from the Floquet unitary, $\Floq$, that evolves the system by one time step. For Hamiltonian or Floquet systems, it is convenient to measure the ratio of consecutive energy gaps (the ``$r$ ratio'') \cite{rratio}; another robust probe of spectral rigidity is the two-point spectral form factor (SFF)  \cite{bertini2018exact,FyodorovMirlin,CDLC1,CDLC2,U1FRUC,ShenkerRMT,ShenkerRMT,SwingleSFFHydro,SwingleSFFSSB,Kos_2021}, 
\be \label{eq:Ktdef}
K(t) \equiv \sum\limits_{m,n = 1}^{\Hdim} \, \overline{\, e^{\mathbf{i} \left( \theta_m - \theta_n \right) t } }\, = \overline{\, \left| \trace [ \, \Floq^t \, ] \right|^2 \,}~~,~~~
\ee
where $\{\theta^{\,}_m\}$ are the eigenphases of $\Floq$, $\Hdim=\Ldim^L$ with $L$ the number of sites and $\Ldim$ states per site, and the overline denotes averaging over an ensemble of statistically similar systems. In the limit $\Ldim \to \infty$, RUCs reproduce the spectral properties of nonlocal random matrix models \cite{BnB,CDLC1}: $K(t) = t$ for $0 < t < \theis = \Hdim$, the Heisenberg time, and $K(t) = \Hdim$ for $t>\theis$. In this limit, thermalization---characterized by a linear ramp $K = t$---is instantaneous.

Away from this limit, one expects an initial overshoot of the linear ramp until interactions thermalize the system \cite{CDLC2,SwingleSFFHydro}. A noninteracting Floquet RUC has $K = t^L$; one can imagine divvying the system into weakly interacting 
blocks of size $\xi (t)$, so that $K(t) \sim t^{L/\xi (t)}$, with $\xi (0) \sim 1$. Under time evolution, interactions lead $\xi (t)$ to grow, saturating to $\xi  (t) = L$ for $t \geq \tth$, so that $K(t) = t$ \cite{ShenkerRMT,CDLC1,ProsenRMTChaosPRX,bertini2018exact,U1FRUC}. The Thouless time, $\tth$---in analogy to single-particle disordered wires \cite{Thouless74,ThoulessThinWire}---is the time it takes for a chaotic system to thermalize fully, signaled by a linear ramp, $K(t) = t$. 

One can also observe delayed thermalization even for $\Ldim \to \infty$ with conserved charges \cite{ShenkerRMT,U1FRUC}. Symmetries (and constraints) lead to independent sectors of $\Floq$ whose eigenvalues do not repel; thus, a chaotic system with $\numsec$ independent sectors will have $K(t) = \numsec \, t$ after thermalizing \cite{U1FRUC}. Ref.~\citenum{U1FRUC} provides a recipe for computing the SFF in the presence of symmetries, mapping $K(t)$ to a classical Markov process, itself equivalent to a quantum Hamiltonian at an RK point~\cite{RK,U1FRUC,SanjayAmosFracton}. Study of the corresponding classical lattice gas reveals that diffusion of the $U(1)$ conserved charge delays thermalization, with $K(t) \to \numsec \, t$ for $t \gtrsim \tth \sim L^2$. Slower, subdiffusive scalings of $\tth$ have also been observed in systems with dipole-moment conservation~\cite{SanjayAmosFracton}.

In this work, we investigate the effect of constraints and symmetries on thermalization by studying the SFF and linear response (connected) transport correlators,
\be \label{eq:Cdef}
\Corr (x,t) \, = \, \expval{\, \charge (x,t) \, \charge (0,0) \, }^{\,}_{c} ~~,~~
\ee
with $\charge (x)$ the local charge density, $\totcharge =\int dx \, \charge (x)$ the conserved $U(1)$ charge, and $\expval{ \, \dots \, } \,  = \, \Hdim^{-1} \, \tr{\, \dots \,}$, the equilibrium average at infinite temperature \footnote{Subtraction of the disconnected part is implied. Connected correlators in the spin and particle language are identical up to a factor of four.}. We provide a recipe for computing the structure factor~\eqref{eq:Cdef} for arbitrary $\Ldim$, and the SFF \eqref{eq:Ktdef} for $\Ldim \to \infty$ in generic, quantum chaotic models, using the machinery of RUCs. We show that the important physics of \emph{both} quantities is controlled by the low energy properties of the \emph{same} transfer matrix, $\Tmat$, which also describes a discrete-time Markov process with the same conservation laws and constraints~\footnote{Such a direct connection between the spectral form factor and two-point correlators was first suggested in Ref.~\citenum{ShenkerRMT}.}. We can also view $\Tmat^{t} \approx e^{- t \, \Ham^{\,}_{\rm RK}}$, where $\Ham^{\,}_{\rm RK}$ lies at an unfrustrated RK point~\cite{RK,U1FRUC,SanjayAmosFracton}. Within a fixed charge sector, the gap of $\Tmat$ (or $\Ham^{\,}_{\rm RK}$) scales as $\gap \sim L^{-z}$; its inverse is the Thouless time, $\tth \sim L^z$, the time required for information to relax throughout the system. The same dynamical exponent controls transport properties from~\eqref{eq:Cdef}, and we find a universal scaling form $ \Corr (x,t) \sim t^{-1/z} f(x/t^{1/z})$, with $z=2$ and $f(\cdot)$ Gaussian for diffusive systems, and $z>2$ for subdiffusive systems.

\begin{figure}[t!]
\centering
	\includegraphics[width = 0.95\columnwidth]{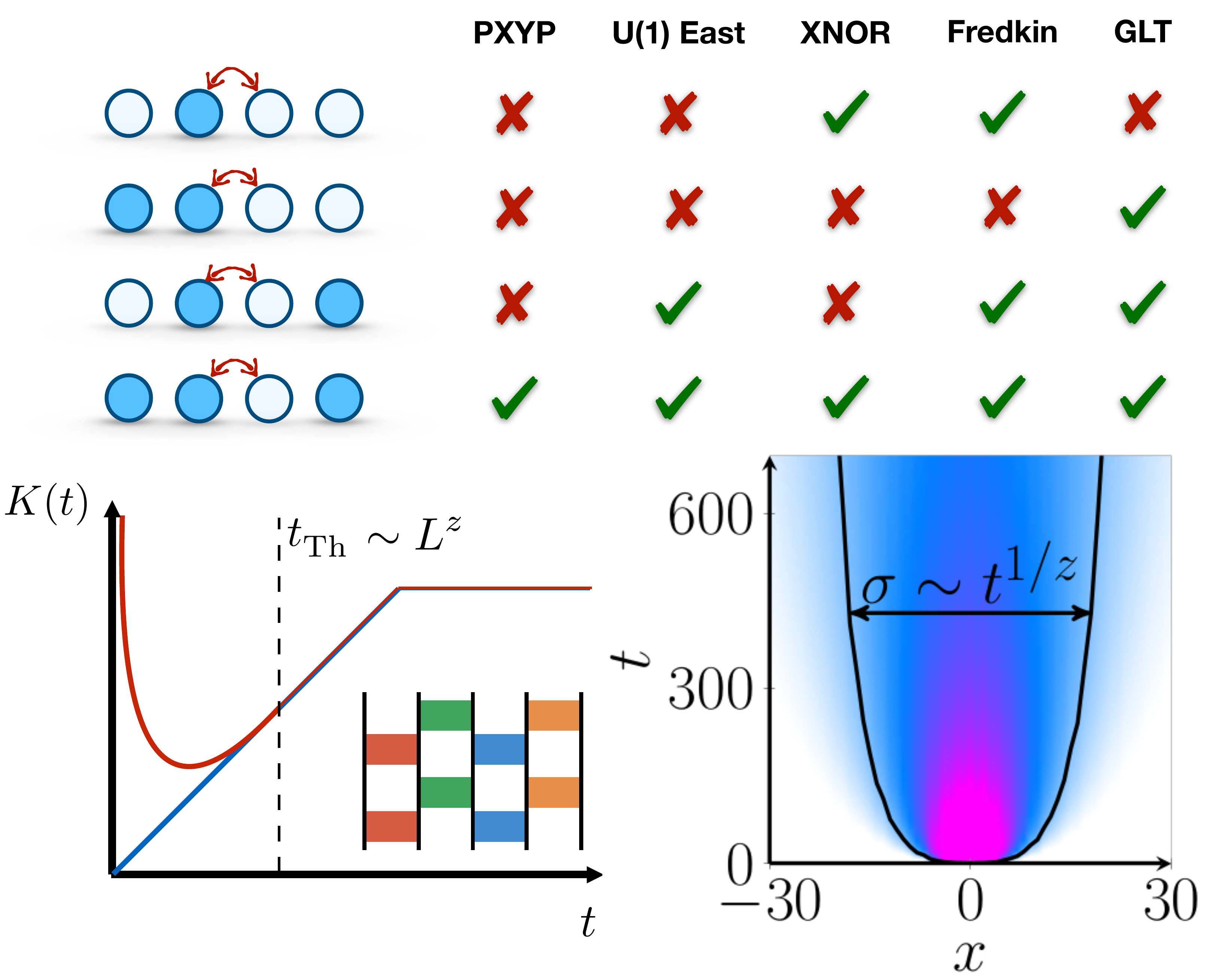} 
	\caption{
	{\bf Models and setup.} {\it Top}:~Cartoon depiction of the allowed dynamical moves for the five models presented; $\particle$ indicates a particle and $\hole$ denotes a hole. {\it Bottom Left}:~Cartoon sketch of the spectral form factor for a generic chaotic system (red) and the RMT prediction (blue); the linear ramp regime ($K(t) = \numsec t$) sets in for $t \gtrsim t^{\,}_{\rm Th}$, the Thouless time, which scales as $L^z$. {\it Bottom Right}:~Heat map of the structure factor (charge two-point function), shown here for Fredkin RUCs, with the variance used to extract $z$, the dynamical critical exponent ($z \simeq 8/3$ for Fredkin constraints).
	\label{fig1} }
\end{figure}

\noindent \emph{Models.---}\, We consider several constrained models acting on a chain of $L$ qubits ($\Ldim=2$) which may be occupied ($\particle$) or empty ($\hole$), with a $U(1)$ conserved charge corresponding to particle number. A cartoon of the allowed dynamical moves is given in Fig.~\ref{fig1}: Essentially, particles are allowed to hop if the neighboring sites are appropriately occupied/unoccupied. Time evolution is generated by a circuit of gates with the general form
\be \label{eq:GateForm} \gate^{\,}_r \, = \, \sum\limits_{\alpha} \, P^{\,}_{r,\alpha} \, \Haar^{\,}_{r,\alpha} \, P^{\,}_{r,\alpha} \, + \, \sum\limits_{\beta} \, P^{\,}_{r,\beta}~~,~~\ee
where $\alpha$ labels constraint-satisfying configurations of cluster $r$ with fixed $U(1)$ charge $\totcharge^{\,}_r = \sum_{j\, \in \, r} \, \charge^{\,}_j $; $\beta$ labels individual constraint-violating configurations on $r$ (with no corresponding unitary dynamics); and $\Haar^{\,}_{r,\alpha}$ is a $n^{\,}_{\alpha} \times n^{\,}_{\alpha}$ Haar unitary that mixes the $n^{\,}_{\alpha}$ states in block $\alpha$ with a fixed $U(1)$ charge \cite{RUCconTibor,RUCconVedika,U1FRUC}. 

The allowed moves for the models considered are depicted in Fig. \ref{fig1}. The Fredkin model \cite{FredkinEntanglement,Korepin2016fredkin,Korepin2017fredkin,ChenFredkin2017,ChenFredkin2017a,Zhang2017fredkin,Udagawa2017fredkin,FredkinFragmentScar,ChenFredkin2020}, allows hopping between sites $j$ and $j+1$ if $j+2$ is occupied 
\emph{or} $j-1$ is unoccupied
, respectively implemented by gates $\gate^{\,}_{r,R}$ (right) and $\gate^{\,}_{r,L}$ (left). 
The Gon\c{c}alves-Landim-Toninelli (GLT) model~\cite{GLT} allows hopping if \emph{either} neighboring site is occupied; XNOR allows hopping if both neighboring sites are in the \emph{same} state~\cite{PhysRevLett.124.207602,10.21468/SciPostPhysCore.4.2.010,PhysRevLett.125.245303,2021arXiv210502252P,2021arXiv210600696P,2021arXiv210804845B}; $U(1)$ East allows hopping only if the right (``East'') neighbor is occupied; and PXYP \footnote{The PXYP model is a $U(1)$ conserving version of the PXP model describing Rydberg atom chains.} allows hopping only if \emph{both} neighboring sites are occupied \cite{KCMRydberg,Bernien2017,Turner2018,PhysRevLett.121.085701,MotrunichRydbergScar,Bluvstein_2021}. Each model is implemented via minimal gates of the form given in Eq.~\ref{eq:GateForm}
. Each type of $\ell$-site gate,  $\gate^{\,}_r$, requires $\ell$ layers per ``time step'', and is always block diagonal in the charge basis~\cite{supp}. Models with different constraints 
or encodings thereof are also discussed in the Supplement~\cite{supp}.

\noindent \emph{Spectral form factor.---}\, Evaluating the SFF \eqref{eq:Ktdef} requires the use of Floquet circuits to guarantee a spectrum: The unitaries comprising the first time step, $\Floq$, are drawn independently; evolution to time $t$ is generated by $\Floq^{\, t\,}$. For arbitrary $t$, ensemble averaging Eq.~\ref{eq:Ktdef} is generally intractable \cite{YoshidaCBD,CDLC1,U1FRUC,SanjayAmosFracton}; to simplify Haar averaging---and to wash out any features not related to particular symmetries and constraints---we include an ancillary $\ancilla$-dimensional qudit on each site \footnote{Note that including the ancillary qudit eliminates any $\beta$-type blocks from Eq.~\ref{eq:GateForm}, as all blocks act nontrivially on the qudits.} so $\Ldim=2 \ancilla$, and take the limit $\ancilla \to \infty$ \cite{U1FRUC,SanjayAmosFracton}. The leading contribution to $K(t)$ can be evaluated diagrammatically~\cite{BnB}, and yields $t$ equivalent ``Gaussian'' diagrams \cite{BnB,CDLC1,U1FRUC}. This procedure is fully generic \cite{U1FRUC,SanjayAmosFracton,supp}: The Haar averaging contracts the indices of gates in the two traces, eliminating one trace as well as the $\ancilla$-state variables, leaving only a single trace over the physical qubits,
\be \label{eq:KtAvgd} K(t) \, = \, t \, \tr{ \, \Tmat^{\, t \,}\,}~~,~~\ee
where the transfer matrix, $\Tmat$, encodes the contribution of configurations of the physical qubits to $K(t)$. 

The form of $\Tmat$ for such models is simple \cite{U1FRUC,SanjayAmosFracton,supp}: $\Tmat$ is a circuit with the same geometry as $\Floq$, comprising Hermitian \footnote{Hermitian gates are to be expected after averaging over a unitary and its conjugate.} gates, $\Tgate^{\,}_r$, i.e. $\Tmat \, = \, \bigotimes_{\lambda} \, \bigotimes_{r \in \lambda} \, \Tgate^{\,}_r $, where $\lambda$ labels layers of the circuit, and  $\Tgate^{\,}_r$, has the same block structure as the corresponding $\gate^{\,}_r$; each block has uniform entries $1/n$, with $n$ the block size \cite{U1FRUC,SanjayAmosFracton,supp}: 
\be \label{eq:Tgate} \Tgate^{\,}_r \, = \, \sum\limits_{\alpha} \,\frac{1}{n^{\,}_{\alpha}} \, \sum\limits_{m,m' \, \in\, \alpha \, } \, \ket{m}\bra{m'}~~,~~\ee
where $m,m'$ run over the $n^{\,}_{\alpha}$ configurations in block $\alpha$. Note that $\Tmat$ describes a discrete-time Markov process for a classical lattice gas with \emph{the same} constraints and conservation laws as the quantum circuit \cite{Schutz2001,U1FRUC,SanjayAmosFracton,supp}. Relatedly, we can define local Hamiltonian terms, $\Ham^{\,}_r = \ident^{\,}_r - \Tgate^{\,}_r$, so that at long wavelengths, $\Tmat^{\, t \,} \approx e^{- t \, \Ham^{\,}_{\rm RK}}$, where $\Ham^{\,}_{\rm RK} = \sum_r \Ham^{\,}_r$ always lies at an unfrustrated RK point \cite{RK,U1FRUC,SanjayAmosFracton}. The Thouless time, $\tth$, marks the start of the linear ramp regime, $K(t) = \numsec \, t$. Each of the $\numsec$ sectors has largest eigenvalue unity; the linear ramp sets in after subleading contributions have decayed: i.e., $\tr{ \Tmat^{\, t \,} } \approx \tr{ e^{- t \, \Ham^{\,}_{\rm RK} }} \,  \to  \, \numsec \left( 1 + e^{- t \, \gap } \right)$ at late times, where $\gap = L^{-z}$ is the gap of  $\Ham^{\,}_{\rm RK}$ (or $\Tmat$). We extract the Thouless time from $K(t)$ as \cite{ShenkerRMT,U1FRUC,SanjayAmosFracton}
\be \label{eq:ThoulessTime} K(t) \sim \, t \,  \left( \numsec  + e^{- t \, \gap} + \dots  \right)~\implies ~ \tth = \frac{1}{\gap} = L^z~,~~~~\ee
with $z$ the dynamical exponent. Thus, $\tth$ gives the time scale over which $K(t) \to \numsec \, t$, and lower bounds the time required for generic models with the same symmetries and constraints to thermalize \cite{SwingleSFFHydro,CDLC1,U1FRUC,SanjayAmosFracton,supp}. For some of the models considered, the low-energy properties of 
$\Ham^{\,}_{\rm RK}$ have been reported: The Fredkin Hamiltonian, e.g., has a gap $\Delta \sim L^{-z}$ with $z>2$~\cite{ChenFredkin2017,ChenFredkin2017a}. Our results imply that the same dynamical exponent, $z$, also controls thermalization and transport properties (see below), for {\em generic} many-body quantum systems 
in this class~\cite{ShenkerRMT}. 

\begin{figure*}[!t]
    \centering
    \includegraphics[width=.95\textwidth]{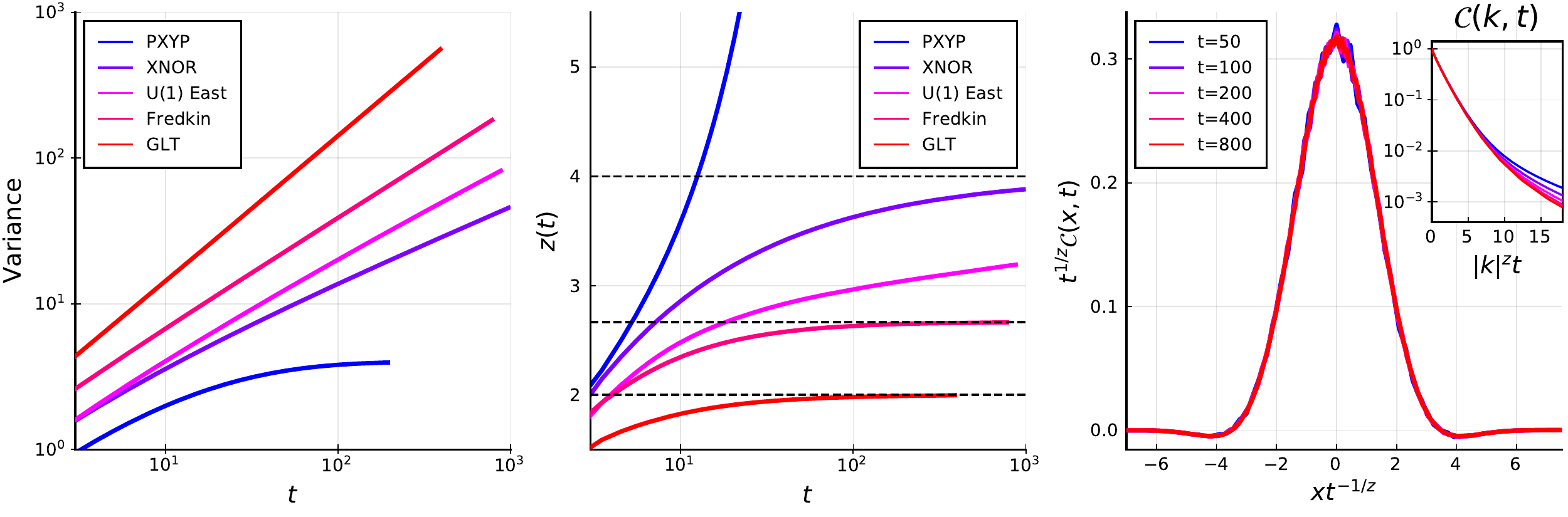} 
    \caption{{\bf Numerical results from the transfer matrix, $\Tmat$. }{\it Left}:~Variance of the spin profile in each of the five models. The variance saturates in the PXYP model, indicating localization, while charges eventually spread across the system in the other models. {\it Middle}:~Apparent dynamical exponent, $z$, versus time, $t$. For GLT,  XNOR and Fredkin, $z(t)$ saturates to $z=2$ (diffusion), $z=4$ and $z=8/3$ (subdiffusion), respectively. In the $U(1)$ East model, $z(t)$ appears to grow without bound,  $z(t) \sim \log t$, indicating quasilocalized dynamics with spread slower than any power law --- another way to extract $z(t)$ using the return probability $\Corr (x=0,t) \sim t^{-1/z}$ leads to a different and even larger estimate $z \approx 7$ for the time scales we can access~\cite{supp}.   {\it Right}:~Collapse of charge profiles for Fredkin when rescaled by the dynamical exponent $z=8/3$. {\it Inset}: Collapse in momentum space showing  ${\cal C}(k,t)\sim {\rm e}^{- C |k|^{z} t}$ at small $k$. TEBD data use maximum bond dimension $\chi^{\,}_{\rm max} = 1024$ for Fredkin and $\chi^{\,}_{\rm max} = 512$ for the other models to ensure convergence.
    \label{fig2} }
\end{figure*}

\noindent \emph{Two-point correlations.---}\, To compute Haar-averaged two-point functions, we dispense with the ancillary qudit and Floquet structure: The five models considered act on $L$ qubits ($\Ldim=2$) with Haar unitaries independently drawn at each time step. Correlators \eqref{eq:Cdef} in RUCs can generically be written in terms of a transfer matrix, 
\be \label{eq:Cij} \Corr^{\,}_{i,j} = \expval{ \,\overline{\, \observ^{\,}_i (t) \, \observ^{\,}_j (0) \, }\,} = \opmatel{\observ^{\,}_i}{\, \Tmat^{\, t \,}\, }{\observ^{\,}_j}~~,~~\ee
where $\opket{\observ}$ is an element of the $\Ldim^{2L}$-dimensional operator space, and $\Tmat$ acts therein, is implicitly Haar averaged \footnote{Each layer of the transfer matrix can be ensemble averaged independently, as unitary gates are independently drawn at each time step.}, and has the same circuit structure as $\Floq$ (and the SFF transfer matrix). For models with Hilbert space dimension $\Ldim$ and unitaries given by Eq.~\ref{eq:GateForm}, the gates of $\Tmat$ take the form~\cite{supp} $\opmatel{\sigma^{\mu}_r}{\Tgate^{\,}_r }{\sigma^{\nu}_r} \, = \, \Ldim^{-\ell} \, \tr{ \, \sigma^{\mu}_r \, \overline{\, \gate^{\dagger}_{r} \, \sigma^{\nu}_r \gate^{\vpd}_r \,}\,}$,
 where $ \opinprod{\sigma^{\mu}_r}{\sigma^{\nu}_r} = \delta^{\,}_{\mu,\nu}$ are orthonormal basis operators (e.g. Pauli strings for $q=2$). Haar averaging gives 
\be \label{eq:CorrTGate} \Tgate^{\,}_r  \, = \, \sum\limits_{\alpha} \, \frac{1}{n^{\,}_{\alpha}} \, \sum\limits_{m,m' \, \in \, \alpha \,} \, | \, \pi^{\,}_m )( \pi^{\,}_{m'} |~~,\ee 
for diagonal (i.e., charge conserving) operators, where $\opket{\pi^{\,}_m} \, = \, \sqrt{\Ldim}\, \ket{m} \bra{m}$ is a projector onto state $m$ in block $\alpha$, and $\opinprod{\pi^{\,}_m}{\pi^{\,}_n} = \delta^{\,}_{m,n}$ form an orthonormal basis for the $\Ldim$ diagonal operators on each site \cite{supp}. 

Crucially, we note that $\Tmat$ \eqref{eq:CorrTGate} is identical to the SFF transfer matrix \eqref{eq:Tgate}, with the $\Ldim$ states per site replaced by $\Ldim$ charge-conserving operators. Thus, the universal features of both spectral and physical correlations are controlled by the low energy spectrum of $\Tmat$, generically relating the Thouless time (related to spectral rigidity) to transport properties, as proposed in Ref.~\citenum{ShenkerRMT}. 
As an aside, we note that nondiagonal (charge-changing) operators do not mix with diagonal operators under $\Tmat$, but evolve under a different transfer matrix if at all \cite{supp}. 
However, we need only consider correlators of diagonal (charge) operators to extract universal transport properties.

\noindent \emph{Numerics.---}\, 
We efficiently simulate the dynamics generated by $\Tmat$ using time evolving block decimation (TEBD) applied to matrix-product operators (MPOs)~\cite{PhysRevLett.91.147902,PhysRevLett.93.207204,PhysRevLett.93.040502,SCHOLLWOCK201196}, exploiting slow entanglement growth compared to the underlying unitary dynamics. We simulate infinite-temperature correlation functions $\Corr (x,t) = \, \Hdim^{-1} \, \tr{ \, \overline{ \, \charge (x, t) \, \charge (0,0) \,} \,}$, where $\charge (x,t)$ is the occupation of site $x$ at time $t$. 
We also use the spatial variance of 
the correlator,~$\sigma^{2}(t) = \sum_{x} \, x^2 \, \Corr(x,t) - (\sum_{x} \, x \, \Corr(x,t))^{2} \sim t^{2/z}$, to extract
the dynamical exponent, $z$---characterizing the transport of charge---via 
$2/z(t) \equiv   d \log \sigma^2 / d \log t $ (shown in the center panel of Fig.~\ref{fig2}).
For GLT, $z(t) \to 2$, indicating diffusive transport and consistent with classical results \cite{GLT}; for XNOR and Fredkin, $z=4$ and $z \simeq 8/3$, respectively, indicating subdiffusion; 
for PXYP, $\sigma^{2}(t)$ itself saturates, indicating localization; and for $U(1)$ East, $z(t)$ grows slowly without saturation, indicating 
quasilocalization.

The PXYP and $U(1)$ East cases can be understood in terms of {\em Hilbert space fragmentation} \cite{RahulFractonCircuit,RahulShattering,PabloFragmentPRX,moudgalya2019thermalization,SLIOM1,kohlert2021experimental,SLIOM1}: The number of sectors, $\numsec$, for both models scales exponentially in system size~\cite{supp}. In the terminology of Ref.~\citenum{PabloFragmentPRX}, PXYP is ``strongly fragmented'' and does not thermalize (i.e. there is no transport; charges are localized), while $U(1)$ East is ``weakly fragmented'' and thermalizes very slowly ($\sigma^2(t)$ grows more slowly with $t$ than any power law). XNOR also shows weak fragmentation, and its dynamical exponent, $z=4$, can be derived analytically from the unusual spin-wave spectrum, $E(k) \sim k^2/L^2$ of the underlying effective RK Hamiltonian~\cite{supp}. This subdiffusive transport with $z=4$ can also be understood in terms of the ``screening'' of the effective charge carried by the diffusive magnon excitations in this model~\cite{PhysRevLett.122.127202,supp}.

We remark that $z(t)$ appears to approach $8/3$ for Fredkin constraints---a similar numerical estimate, $z \approx 2.69$, was reported in Ref.~\citenum{ChenFredkin2017a} in the context of low-temperature physics of the Fredkin Hamiltonian. While our results derive from RUCs, we expect that this $z=8/3$ characterizes a new dynamical universality class of \emph{generic} many-body quantum or classical systems (Floquet, Hamiltonian, or noisy) with Fredkin constraints. The Fredkin correlator satisfies the universal scaling form $\Corr(x,t) \sim 1/t^{1/z} f(x/t^{1/z})$, with $f(\cdot)$ a non-Gaussian function (see third panel of Fig.~\ref{fig2} and Ref.~\citenum{supp}).  Remarkably, we find numerically that the Fredkin RK Hamiltonian has a low-energy spectrum $E(k) \sim k^{4/3}/L^{4/3}$, reminiscent of the XNOR model, suggesting the possibility of a similar mechanism for subdiffusion in both models~\cite{supp}.

\noindent \emph{Discussion.---}\, 
We studied spectral and transport properties of many-body quantum systems with conserved charges and kinetic constraints using random unitary circuits. We computed ensemble-averaged spectral form factors and linear-response correlation functions for various classes of constraints, and showed that both relate to the same transfer matrix, $\Tmat$, describing a classical Markov process
. This mapping holds for \emph{any} choice of symmetries and constraints, and in any dimension; however, numerical simulation of $\Tmat$ becomes intractable for $d>1$. Our results establish a general correspondence between the Thouless time and transport properties in conserving systems, and we unearth a broad range of possible transport properties depending on the 
constraint. The Fredkin universality class is especially interesting: Further characterizing its dynamical exponent of $z \simeq 8/3$ presents a clear direction for future work.

\noindent \emph{Acknowledgments.---}\, 
We thank U. Agrawal, J. T. Chalker, A. De Luca, and A.C. Potter for useful discussions and collaborations on related work; we thank J. P. Garrahan, S. Gopalakrishnan, A. Lucas, R. Nandkishore, B. Pozsgay, T. Rakovszky, P. Sala, and W. Witczak-Krempa for the same and for their feedback on this manuscript. We acknowledge support from the Air  Force  Office  of  Scientific  Research  under  Grant No. FA9550-21-1-0123  (RV and BAW) and the Alfred P. Sloan Foundation through a Sloan Research Fellowship (RV).

\noindent {\it Note Added.---}\, While completing this manuscript, Ref.~\citenum{Pal2021Motzkin} appeared on the arXiv, and reports subdiffusive hydrodynamics for the ``Motzkin'' Hamiltonian; Motzkin constraints are very similar to Fredkin constraints, and appear to lie in the same universality class with dynamical exponent $z\simeq 8/3$~\cite{supp}.

\bibliography{RUFC}

\bigskip

\onecolumngrid
\newpage



\section*{Supplementary material}

# Supplemental Material: Subdiffusion and many-body quantum chaos with kinetic constraints

Hansveer Singh,[1] Brayden A. Ware,[1] Romain Vasseur,[1] and Aaron J. Friedman[2]

[1]*Department of Physics, University of Massachusetts, Amherst, Massachusetts 01003, USA*
[2]*Department of Physics and Center for Theory of Quantum Matter,
University of Colorado, Boulder CO 80309, USA*

## CONTENTS

## 1. EVOLUTION WITH BLOCK STRUCTURE

We consider models acting on chains of $L$ sites with a $q$-state Hilbert space on each site. Dynamics are generated by applying random unitary circuits (RUCs) [1–5] comprising $\ell$-site gates, $\mathcal{U}_r$, where $r$ labels distinct $\ell$-site clusters. Specific models may require the application of multiple types of gates (e.g., for the Fredkin model, where hopping is allowed if the right site is occupied *or* the left site is unoccupied). A single time step requires $\ell$ layers of each gate type, tiled in a brick-wall geometry to cover all sites. For two-site gates, the two layers correspond to even and odd bonds.

Each unitary gate can be written as

$$\mathcal{U}_r = \sum_\alpha P_r^\alpha \, U_{r,\alpha} \, P_r^\alpha + \sum_\beta P_r^\beta \quad , \tag{1.1}$$

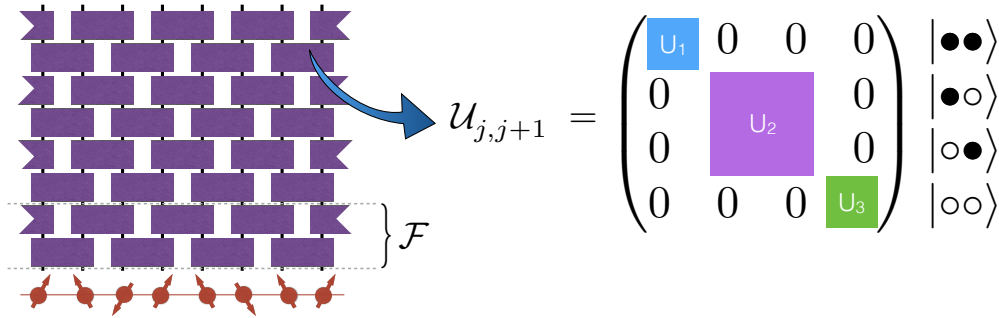

FIG. 1. **Charge conserving two-site gate.** The block structure corresponds to different $U(1)$ charges (i.e. number of particles or $z$ component of spin), which are conserved by each gate: The three blocks correspond to $0, 1, 2$ particles (or $S^z_{\text{tot}} = -1, 0, +1$), and unitary gates do not mix between blocks. Rather, each block contains an independently drawn Haar random unitary. The $1 \times 1$ blocks are simply complex phases, unless we tensor in an additional $d$-dimensional qudit, in which case we allow $d^2 \times d^2$ Haar unitaries acting on the ancilla qudits (with a $2d^2 \times 2d^2$ gate acting in the nontrivial block).

where $\alpha$ labels blocks of states that are allowed to mix together (i.e., configurations of cluster $r$ with fixed charge, $Q_r = \sum_{j \in r} \mathfrak{q}_j$ and that satisfy any constraint), $\beta$ labels projectors onto individual states with no dynamics (and hence, the "unitary" is simply the identity), and $U$ always denotes a Haar random unitary matrix [1, 6] that acts in the subspace of states in block $\alpha$. Unitarity of $\mathcal{U}_r$ requires

$$\sum_\alpha P_r^\alpha + \sum_\beta P_r^\beta = \mathbb{1} \; .$$

We take the projectors $P_r^\beta$ to project onto a single $\ell$-site configuration, for simplicity: These projectors arise in constrained circuits only, corresponding to configurations that do not satisfy the constraint, and therefore dynamical moves are forbidden [1]. Here, the term "block" refers to specific sectors of states that are permitted to mix, so the block size corresponding to any configuration that does not satisfy a constraint is one, as these states do not mix with other configurations.

The number of blocks, $N$, is strictly less than $q^\ell$: In the case of conserved charges, $N$ is the number of different total charges that the $\ell$-site cluster can realize; in the case of constraints, $N$ simply reflects the number of unique dynamical moves allowed on the cluster.

Each Haar gate, $U_{r,\alpha}$, in Eq. 1.1 is an $n_\alpha \times n_\alpha$ random unitary acting on the $n_\alpha$ configurations of the $r$th cluster that are not annihilated by $P_r^\alpha$. For concreteness, we can write

$$P_r^\alpha = \sum_{m_1,\ldots,m_\ell=1}^{q} c^\alpha_{m_1,\ldots,m_\ell} P_j^{(m_1)} \ldots P_{j+\ell}^{(m_\ell)} \quad \text{with} \quad P_j^{(m)} = |m\rangle\langle m|_j \; , \tag{1.2}$$

i.e. $P_j^{(m)}$ are the "naïve" projectors that project site $j$ into state $|m\rangle$.

The block size, $n_\alpha$, is simply the number of $\ell$-site states that are not killed by the projector $P_r^\alpha$, which is also the number of distinct $\ell$-site *naïve* projectors that make up $P_r^\alpha$. Note that the projectors $P_j^{(\beta)}$ are naïve by construction, as they project onto a single configuration.

As a concrete example, Fig. 1 depicts a spatial slice of a $U(1)$ circuit acting on qubits ($q = 2$) [7–10]. It is sufficient to use two-site gates, $\mathcal{U}_{j,j+1}$, with layers alternating between even/odd $j$. Each gate conserves the number of particles living on the pair of sites upon which it acts, i.e. $Q_{j,j+1} = n_j + n_{j+1}$ (or in the

---

[1] In models with symmetries alone, we may allow $1 \times 1$ Haar unitaries to act in single-state blocks to make the evolution "more chaotic".

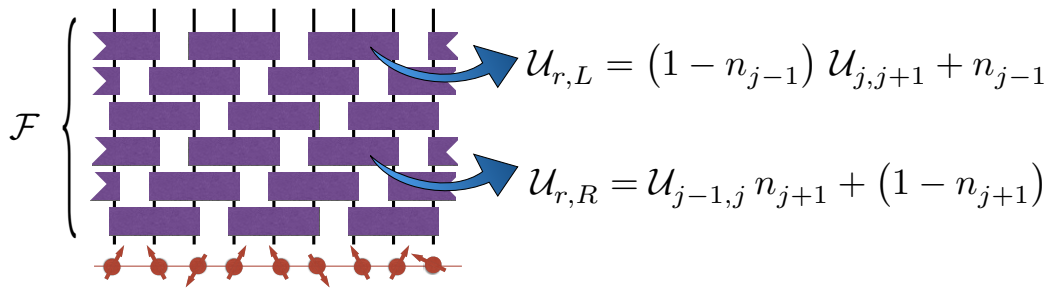

FIG. 2. **Fredkin gates.** There are two distinct three-site gates, labelled $L$ and $R$, corresponding to whether the constraint comes from the left ($L$) or right ($R$) neighbor. If the left neighbor is unoccupied ($\circ$) or the right neighbor is occupied ($\bullet$) then we apply a $U(1)$ gate, $\mathcal{U}_{j,j+1}$, which itself has three blocks corresponding to $0, 1, 2$ particles, as depicted in Fig. 1. Otherwise, we act with the identity. Each site, $j$, should be the left, center, and right site of both the $L$ and $R$ gates once per time step.

spin-1/2 language, the $z$-component of spin is conserved by each gate). Consequently, the gate acquires a block structure, leaving the doubly occupied ($|\bullet\bullet\rangle$) and doubly unoccupied ($|\circ\circ\rangle$) states unchanged, while allowing hopping in the case of only one particle on the two-site cluster ($|\bullet\circ\rangle \leftrightarrow |\circ\bullet\rangle$).

As an example with both particle conservation and kinetic constraints, we depict the Fredkin circuit in Fig. 2. Again, the model is defined on qubits, which can be occupied ($\bullet$) or empty ($\circ$), and we use two types of three-site gates: The "$R$" gate applies a $U(1)$ conserving gate to sites $j-1$ and $j$ if the right ($R$) site, $j+2$ is occupied ($\bullet$); the "$L$" gate applies a $U(1)$ gate to $j$ and $j+1$ if the left ($L$) site ($j-1$) is unoccupied ($\circ$); if the constraint is not satisfied, no Haar unitary is applied (we act with the identity). As with the unconstrained $U(1)$ circuit, each $U(1)$ gate within a Fredkin gate is itself block diagonal, with independently drawn Haar unitaries acting within each fixed-charge block. The Fredkin circuit has six layers per "time step": site $j$ is the left, middle, and right site of both $L$ and $R$ type gates exactly once every time step.

Other models with any combination of constraints and conservation laws can be written down using a combination of [mutually commuting] projectors and Haar random unitaries, of the form given by Eq. 1.1. The same model may be written in different ways by using different size gates. For the models considered, our choices of constraint encoding does not appear to affect the chaotic properties of the system, although this can happen, e.g. in dipole-conserving circuits [11]. The recipe for computing the transfer matrices corresponding to the spectral form factor and correlation functions is the same for any model defined in this way.

## 2. HAAR AVERAGED SPECTRAL FORM FACTOR

The two-point spectral form factor (SFF) is the Fourier transform of the two-point correlator of eigenvalues of a Hamiltonian, $H$ (or Floquet unitary, $\mathcal{F}$) [5, 9, 12–15]. If a system is thermal, then it forgets everything it possibly can, excepting any information related to conserved quantities (or, more generally, each independent sector of the evolution operator is expected to relax internally due to spectral rigidity). Thus, a thermal system should have the same properties as one evolved by a random matrix with the same symmetries (or independent sectors).

The SFF for such chaotic random matrix models takes the form [5, 12, 15]

$$K(t) = \begin{cases} \mathcal{D}^2 & t = 0 \\ \mathcal{N} t & 0 < t < t_{\text{Heis}} \\ \mathcal{D} & t > t_{\text{Heis}} \end{cases}, \qquad (2.1)$$

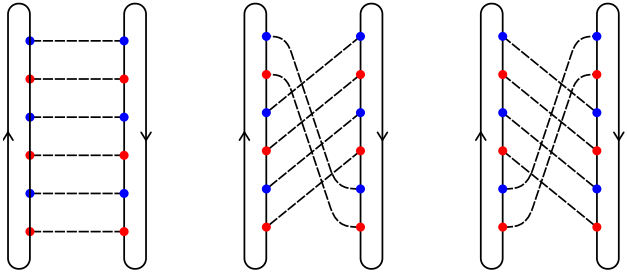

FIG. 3. **Gaussian diagrams.** Example of the contractions of unitaries and their conjugates as seen in Refs. 5, 9, and 10. The above correspond to cyclic "Gaussian" diagrams, the leading contribution to the spectral form factor calculation. The above depicts only the unitaries acting on a single cluster, $r$—the contractions of all other unitaries in the circuit inherit from the first contraction due to the "bond constraint".

where $t_{\text{Heis}} = \mathcal{D}$ is the inverse mean level spacing, $\mathcal{D}$ is the many-body Hilbert space dimension, and $\mathcal{N}$ is the number of independent sectors. If one looks only within a single sector, one expects a linear ramp $K(t) = t$, and $\mathcal{D}$ should refer to the dimension of the sector.

The spectrum need not correspond to a Hamiltonian (i.e. continuous time translation symmetry is not required): The quantity is well defined for Floquet models, where the "spectrum" refers to the eigenphases, $\{\theta_n\}$, of the Floquet evolution operator, $\mathcal{F}$, that evolves the system by a single time step. The SFF for Floquet RUCs can be written as

$$K(t) \equiv \sum_{m,n=1}^{\mathcal{D}} \overline{e^{\mathbf{i}(\theta_m - \theta_n)t}} = \overline{|\text{tr}[\mathcal{F}^t]|^2} \quad , \tag{2.2}$$

where the overline denotes Haar averaging (for Hamiltonian models, we replace $\mathcal{F}^t$ with $\mathcal{W}(t) = e^{-\mathbf{i}tH}$, but the formula for $K(t)$ is unchanged).

### 2.1. Haar averaging

To facilitate Haar averaging, we include an ancillary $d$-dimensional qudit in addition to the qubits encoding the conservation law and constraints, so $q = 2d$. Each block in Eq. 1.1, therefore, contains an $n_\alpha d^\ell \times n_\alpha d^\ell$ independently drawn Haar random unitary, where $\ell$ is the number of sites acted on by a single gate, and $n_\alpha$ is the number of [qubit] states in block $\alpha$. As a result, there are no terms in the sum over $\beta$ in Eq. 1.1, further simplifying the Haar averaging when we take the limit $d \to \infty$. While this limit may appear extreme on its face, (i) it is known that one cannot observe certain features, such as the plateau for $t \geq t_{\text{Heis}}$, by including subleading terms (in $1/d$) [5] and (ii) results for $q, L, t \to \infty$ are nonetheless in good agreement with small system and short time numerics with $q = 2$ and $L \sim 12$ [9]. In fact, the large $d$ limit essentially removes any features in $K(t)$ not related to the symmetries and constraints we impose [9].

Returning to Eq. 2.2, we are confronted with a $t$-fold Haar channel [1]. Averaging over the Circular Unitary Ensemble (CUE) with Haar measure fixes the indices of a unitary, $U$, to match those of its conjugate, $U^*$, summing over all pairings of unitaries with their conjugates. For finite $q$, this problem is impossibly difficult for arbitrary $t$, as there are simply too many combinations to account for. However, as $d \to \infty$, only a finite set of pairings contribute, with subleading corrections suppressed by at least $1/d^2$.

The Haar averaging procedure was first explained in the context of FRUCs in Ref. 5, using a diagrammatic method for averaging over the CUE [6], albeit without symmetries or constraints. Ref. 9 details how the averaging works out when a $U(1)$ symmetry is encoded via qubits, and Ref. 10 clarifies that this procedure works more generally beyond the $U(1)$ case, and provides a general prescription. We refer the reader to these works for further detail.

Because the Haar average fixes the indices of every $U^\dagger$ to match those of a $U$ (corresponding to the same gate, $\mathcal{U}_r$ and block, $\alpha$), the averaging will essentially eliminate one of the two traces in Eq. 2.2. The leading

4

diagrams are "Gaussian" [5, 6], with the simplest corresponding to matching some $U_{r,\alpha}$ with its conjugate *at the same time step* (i.e., the same layer of $\mathcal{F}^t$). The other Gaussian diagrams correspond to cyclic shifts of the pairing, so that a Haar unitary in layer $s$ is paired instead with its conjugate in layer $s+\tau$. Because of cyclic invariance of the trace, the labelling of layers is modulo $t$, and there are $t$ such "shifts" of the simple Gaussian diagram, all of which are equivalent to one another by cyclic invariance of the trace. These diagrams are depicted in Fig. 3.

The "bond constraint" [5], which comes from the geometry of the circuit, requires that after we match a gate, $U_{r,\alpha}$, in layer $s$ to $U_{r,\alpha}^\dagger$ in layer $s+\tau$, *all other unitaries* must be paired with their conjugates shifted by $\tau$ layers. Any deviation from this pairing would necessarily be subleading, and can therefore be ignored. The result is $t$ copies of the simplest Gaussian diagram. However, we still have to deal with the block structure of the gates (and overall circuit structure).

Since the unitaries in each block are independently drawn, if we find ourselves in block $\alpha$ for gate $r$ in the $\mathrm{tr}\,[\mathcal{F}^t]$ term, then the Haar average will be zero unless we pair with a $U^\dagger$ also from block $\alpha$ and cluster $r$. The result of Haar averaging is simply $1/n_\alpha$, the size of the block. Considering the physical qubits, if we are in a block with a single qubit configuration, the Haar average requires that the other trace is in that same state, and the corresponding coefficient is one. If we are in a block with several states, then the Haar average only requires that the other trace is in one of the states in the block, but not precisely the same state.

## 2.2. Transfer matrix

We can write down a $2^\ell \times 2^\ell$ matrix, $T_r$, whose first index runs over states corresponding to the $\mathrm{tr}\,[\mathcal{F}^t]$ term, and the second index corresponds to the other trace. This matrix will therefore be block diagonal, with the *same block structure* as $\mathcal{U}_r$ (1.1). In blocks of size one, the result is simply one. In blocks of size $n$, the result is $1/n$ in every entry of the block, as the qubits in the two traces need not be in the same configuration for the Haar average to be nonzero, but simply the same block.

The result is

$$K(t) = t\,\mathrm{tr}\left[\mathcal{T}^t\right]\ , \tag{2.3}$$

where the overall factor of $t$ comes from the $t$ equivalent diagrams, and $\mathcal{T}$ is the transfer matrix, which can also be thought of as effecting a classical Markov process. The transfer matrix has identical geometry to $\mathcal{F}$ in terms of layers and composition of $\ell$-site gates. Each layer consists of Hermitian gates (Hermiticity is due to having averaged unitaries and their conjugates), with $T_r$, given by

$$T_r = \sum_\alpha \frac{1}{n_\alpha} \sum_{m,m'\in\alpha} |m\rangle\langle m'|\ , \tag{2.4}$$

i.e., each block has unit trace, and maps any state in the block to a uniform superposition of all $n_\alpha$ states $\{|m\rangle\}$ in block $\alpha$, with coefficient $1/n_\alpha$.

## 2.3. Thouless time and RK connection

Each gate, $T_r$, has largest eigenvalue one and smallest eigenvalue zero; likewise, $\mathcal{T}$ has largest eigenvalue one and smallest eigenvalue zero *in each independent sector*. Note that each sector is independent, and that only eigenvalues within a sector should show repulsion. The transfer matrix can also be viewed as a discrete "Trotterization" of some Hamiltonian, $H_{\mathrm{RK}}$, with local, $\ell$-site terms $H_r = \mathbb{1}_r - T_r$, where $H_{\mathrm{RK}} = \sum_r H_r$ always lies at a frustration-free Rokhsar Kivelson (RK) point (a connection first reported in Ref. 10). The leading eigenstate, with energy zero, is a uniform superposition of all possible states.

The Thouless time, $t_{\mathrm{Th}}$, is the time at which $K(t) \sim \mathcal{N}t$ (without restriction to a sector). Essentially, there are $\mathcal{N}$ eigenstates of $\mathcal{T}$ with eigenvalue one, and $t_{\mathrm{Th}}$ is the time it takes for all other eigenstates to decay, so that only the steady state of each sector contributes to Eq. 2.3. The time scale of this decay is set by the gap of the transfer matrix, $\mathcal{T}$, between unity and the second largest eigenvalue (within each sector);

equivalently, this is set by the gap of $H_{\text{RK}}$, whose ground state energy is zero [9, 10]. Noting that $\mathcal{T}^t \approx e^{-tH}$ at long wavelengths, we write

$$K(t) \approx t \operatorname{tr}\left[ e^{-tH} \right] \approx t \left( \mathcal{N} + e^{-t/\Delta} + \ldots \right) , \qquad (2.5)$$

where $\Delta(L)$ is the energy of the first excited state (i.e., the gap of $H$ since the ground state energy is $E = 0$), which may vary between sectors, both in terms of scaling with $L$ and prefactors.

Within each sector, $K(t) \to t$ on a time scale $t_{\text{Th}} \sim 1/\Delta$ with $\Delta \sim L^{-z}$, where $z$ is the dynamical exponent. Overall, the approach to the steady state of the Markov process generated by $\mathcal{T}$ is controlled by the *slowest* sector. In the case of a single conserved quantity, one expects $z = 2$ and $t_{\text{Th}} = L^2$, as reported in Ref. 9, and this scaling holds exactly in every sector, due in part to integrability. However, one can also study $\mathcal{T}^t$ directly using matrix product state numerics. We note that $t_{\text{Th}} = L^z$ lower bounds the thermalization time, as indicated by a linear ramp in $K(t)$ [10].

## 3. TRANSFER MATRIX FORMULATION OF TWO POINT CORRELATION FUNCTIONS

A key ingredient of the main text is the ability to compute Haar averaged correlation functions for large systems using a combination of analytics and tensor network numerics. We can write generic two-point correlation functions as

$$\mathcal{C}_{i,j}(t) = \left\langle \overline{\mathcal{O}_i(t) \mathcal{O}_j(0)} \right\rangle_\rho = \operatorname{tr}\left[ \rho \, \overline{\mathcal{W}^\dagger(t) \, \mathcal{O}_i \, \mathcal{W}(t) \, \mathcal{O}_j} \right] , \qquad (3.1)$$

where the overline indicates averaging over an ensemble of statistically similar system. For RUCs, this involves averaging the unitary gates comprising $\mathcal{W}$ over the Circular Unitary Ensemble (i.e., Haar averaging). Since these systems generically heat up to infinite temperature, the corresponding thermal density matrix is simply proportional to the identity, $\rho \propto \mathbb{1}$.

### 3.1. Operator space

We can define an inner product space where the operators acting on the physical Hilbert space of our system live. For a lattice with $L$ sites, each with local Hilbert space dimension $q$, this operator space consists of $\mathcal{D}^2 = q^{2L}$ unique operators. We denote by rounded kets, $|\mathcal{O})$, an element of this operator space corresponding to the observable $\mathcal{O}$, as distinct from states, $|\psi\rangle$, living in the original Hilbert space.

The inner product is defined by

$$(\mathcal{O}|\mathcal{O}') = \frac{1}{\mathcal{D}} \operatorname{tr}\left[ \mathcal{O}^\dagger \mathcal{O}' \right] , \qquad (3.2)$$

and one can also define an orthonormal operator basis given by $\{ \sigma^\mu \}$ [2], satisfying

$$(\sigma^\nu | \sigma^\mu) = \delta_{\mu,\nu} . \qquad (3.3)$$

We can decompose the time dependence of generic operators in this basis,

$$\left| \mathcal{O}_j(t) \right) = \sum_\nu a_\nu^j(t) \, |\sigma^\nu) , \qquad (3.4)$$

where

$$a_\nu^j(t) \equiv \frac{1}{\mathcal{D}} \operatorname{tr}\left[ \sigma^\nu \mathcal{O}_j(t) \right] = \left( \sigma^\nu \, \big| \, \mathcal{O}_j(t) \right) , \qquad (3.5)$$

---

[2] For $q = 2$, the set of Pauli strings, $\sigma^\mu = \bigotimes_j \sigma_j^{\mu_j}$ form such a basis. However, any basis will do.

which is, itself, an infinite-temperature two point correlation function between $\mathcal{O}_j(t)$ and $\sigma^\nu(0)$. The above two equations are implied by completeness of the operator basis: $\sum_\nu |\sigma^\nu)(\sigma^\nu| = 1$.

A general, infinite-temperature correlation function can be written

$$\mathcal{C}_{i,j}(t) = \frac{1}{\mathcal{D}} \text{tr}\left[\overline{\mathcal{O}_i(t)\,\mathcal{O}_j(0)}\right] = \overline{(\mathcal{O}_j(0)|\mathcal{O}_i(t))} = \sum_\nu \overline{a_\nu^{j\,*}(0)\,a_\nu^i(t)} \ . \tag{3.6}$$

### 3.2. Transfer matrix in operator space

The quantity of interest is the ensemble-averaged correlation function, $\mathcal{C}_{i,j}$ (3.6). Let us consider the quantity $a_\mu^i(t)$ corresponding to the time evolved observable, $\mathcal{O}_i(t)$[3],

$$\overline{a_\mu^i(t)} = \frac{1}{\mathcal{D}}\text{tr}\left[\sigma^\mu\,\overline{\mathcal{O}_i(t)}\right] = \mathcal{C}_{i,\mu}(t) \ , \tag{3.7}$$

where the overline denotes Haar averaging.

We can recover an equation governing the dynamics of the Haar averaged coefficients $a_\mu^i(t)$—and therefore, the two-point correlator, $\mathcal{C}_{i,j}(t)$—analytically. We can write the ensemble-averaged coefficient at time $t+1$ in terms of ensemble-averaged coefficients at time $t$ as follows:

$$\overline{a_\mu^i(t+1)} = \frac{1}{\mathcal{D}}\text{tr}\left[\sigma^\mu\,\overline{\mathcal{O}_i(t+1)}\right] = \frac{1}{\mathcal{D}}\text{tr}\left[\sigma^\mu\,\overline{\mathcal{W}_t^\dagger\,\mathcal{O}_i(t)\,\mathcal{W}_t}\right] \ , \tag{3.8}$$

where we have used the fact that the unitary gates at each time step are independently drawn, and their averages may be taken independently. Using Eq. 3.4 we can write $\overline{\mathcal{O}(t)}$ in terms of our operator basis and averaged coefficients, i.e.,

$$= \sum_\nu \overline{a_\nu^i(t)}\,\frac{1}{\mathcal{D}}\text{tr}\left[\sigma^\mu\,\overline{\mathcal{W}_t^\dagger\,\sigma^\nu\,\mathcal{W}_t}\right] = \sum_\nu \mathcal{T}_{\mu,\nu}^{(t)}\,\overline{a_\nu^i(t)} \ , \tag{3.9}$$

where $\mathcal{W}_t$ is the circuit of unitary gates that evolves the system from time $t$ to time $t+1$, and we have implicitly defined the *transfer matrix* for the $t$th time step

$$\mathcal{T}_{\mu,\nu} = (\sigma^\mu|\mathcal{T}|\sigma^\nu) = \frac{1}{\mathcal{D}}\text{tr}\left[\sigma^\mu\,\overline{\mathcal{W}_t^\dagger\,\sigma^\nu\,\mathcal{W}_t}\right] \ , \tag{3.10}$$

where the form of the RHS will depend on the particulars of the circuit.

In practice, a single "time step" will involve the application of two or more layers of untiary gates, reflected in the composition of $\mathcal{W}_t$. In the case of two-site gates, there are two layers of circuit per time step, corresponding to gates acting on even vs. odd bonds, and the transfer matrix for one time step takes the form

$$\mathcal{T} = \mathcal{T}_e \mathcal{T}_o \quad \text{with} \quad \mathcal{T}_{e/o} = \bigotimes_{j\in e/o} T_{j,j+1} \ , \tag{3.11}$$

where $T$ refers to a particular gate in the layer.

The circuits of interest have a fixed geometry; correspondingly, the transfer matrix, $\mathcal{T}$ for one time step will be independent of time, $t$, and will comprise layers of Hermitian, $\ell$-site gates, $T_r$, arranged in precisely the same pattern as the gates $\mathcal{U}_r$ comprising $\mathcal{W}_t$. We denote individual layers of $\mathcal{T}$ by $\mathcal{T}_{(\lambda)}$, and the gates within such layers by $T_r$, where $r$ labels $\ell$-site clusters, and the clustering of sites depends on the layer, $\lambda$.

We can use this transfer matrix (3.10) to write Eq. 3.9 as

$$\overline{a_\mu^i(t)} = \sum_\nu (\sigma^\mu|\mathcal{T}^t|\sigma^\nu)\,\overline{a_\nu^i(0)} \ , \tag{3.12}$$

---

[3] Which itself can be regarded as a correlation function if $\mathcal{O}_j \to \sigma^\mu$.

and noting that $a^i_\mu(t) = (\sigma^\mu | \mathcal{O}_i(t))$, the above is equivalent to

$$\left| \overline{\mathcal{O}(t)} \right) = \mathcal{T}^t \left| \overline{\mathcal{O}(0)} \right) , \tag{3.13}$$

and implies for the infinite temperature two-point correlation function (3.1)

$$\mathcal{C}_{i,j}(t) = \langle \mathcal{O}_i(t) \mathcal{O}_j(0) \rangle = \left( \mathcal{O}_j | \mathcal{T}^t | \mathcal{O}_i \right) , \tag{3.14}$$

where all quantities above are ensemble averaged.

We note that the largest eigenvalue of each layer of $\mathcal{T}$ is always unity, and the smallest always zero. Thus, for large systems and at late times, the correlation function in Eq. 3.14 is dominated by the gap between the second largest eigenvalue and unity. Equivalently, one can consider an effective local Hamiltonian term for the cluster, $r$, given by $H_r = \mathbb{1}_r - T_r$, whose ground state has energy 0, and whose gap controls the long time behavior of Eq. 3.14.

### 3.3. General form of the transfer matrix

Now let us consider the form of the transfer matrix for a generic model with block-diagonal gates. These blocks result from the inclusion of projectors, which may encode conservation laws and/or kinetic constraints. To properly encode *either* constraints or Abelian circuits, all projectors must be in the same basis (i.e., the "$z$" basis); consequently, the block diagonal form of the gates (in the preferred basis) is guaranteed.

For this calculation, we dispense with the "ancillary" $q$-dit used to facilitate Haar averaging in computing the spectral form factor: Any nonidentity operators acting on these $q$-dits will be annihilated after application of any layer of the transfer matrix, $\mathcal{T}$, and their inclusion is therefore unnecessary. Only degrees of freedom upon which projectors act can give rise to nontrivial transfer matrices. In the remainder, $q$ is the physical Hilbert space dimension on each site.

Assuming a circuit with gates of the form given in Eq. 1.1, we now compute layer $\lambda$ of the transfer matrix:

$$\mathcal{T}^{(\lambda)}_{\mu,\nu} = \left( \sigma^\mu \Big| \mathcal{T}_{(\lambda)} \Big| \sigma^\nu \right) = \frac{1}{q^L} \operatorname{tr} \left[ \sigma^\mu \, \overline{\mathcal{W}^\dagger_\lambda \sigma^\nu \mathcal{W}_\lambda} \right] = \prod_{r(\lambda)} \frac{1}{q^\ell} \operatorname{tr} \left[ \sigma^\mu \, \overline{\mathcal{U}^\dagger_r \sigma^\nu \mathcal{U}_r} \right] , \tag{3.15}$$

where $r(\lambda)$ indicates that the clustering of sites depends on the layer, $\lambda$. Using Eq. 1.1 and, noting that the Haar average is zero unless unitaries, $U_{r,\alpha}$, are paired with their conjugates, $U^\dagger_{r,\alpha}$, with the same $\alpha$, we have

$$= \prod_r \frac{1}{q^\ell} \left\{ \sum_\alpha \operatorname{tr} \left[ \sigma^\mu \, P^\alpha_r \, \overline{U^\dagger_{r,\alpha} P^\alpha_r \sigma^\nu P^\alpha_r U_{r,\alpha}} \, P^\alpha_r \right] + \sum_{\beta,\beta'} \operatorname{tr} \left[ \sigma^\mu \, P^\beta_r \, \sigma^\nu \, P^{\beta'}_r \right] \right\} , \tag{3.16}$$

where the projector indices of the first term are required to be equal by Haar averaging, but the second term did not involve a Haar average. Evaluating the Haar average gives

$$\mathcal{T}^{(\lambda)}_{\mu,\nu} = \prod_r \frac{1}{q^\ell} \left\{ \sum_\alpha \frac{1}{n_\alpha} \operatorname{tr} [P^\alpha_r \, \sigma^\mu] \operatorname{tr} [P^\alpha_r \, \sigma^\nu] + \sum_{\beta,\beta'} \operatorname{tr} \left[ \sigma^\mu \, P^\beta_r \, \sigma^\nu \, P^{\beta'}_r \right] \right\} , \tag{3.17}$$

### 3.4. Unitary operator basis.

The Pauli strings form a complete, orthonormal basis of operators acting on $L$ qubits. The Paulis are both Hermitian *and* unitary, and generalizations of the Pauli string basis of both types are possible when we take $q > 2$. For general $q$, it will be convenient to use the unitary—or "Weyl"—basis:

$$\Gamma_{m,n} = X^m Z^n , \text{ where } X \equiv \sum_{n=0}^{q-1} |n+1\rangle\langle n| \text{ and } Z \equiv \sum_{n=0}^{q-1} \omega^n |n\rangle\langle n| , \tag{3.18}$$

where $\omega = \exp(2\pi i/q)$. The operator $X$ is a "shift" (or "cycle") operator, and $Z$ is a "weight" operator, to borrow the terminology of $\mathbb{Z}_q$ clock models, where these operators commonly appear. Also note:

$$|n + q\rangle \cong |n\rangle \,, \quad X^q = Z^q = \mathbb{1} \,, \quad \text{tr}\,[X] = \text{tr}\,[Z] = 0 \,, \quad \text{and } \omega XZ = ZX \,.$$

Using these conventions we can write

$$\Gamma_{m,n} = \sum_{k=0}^{q-1} \omega^{kn} |k+m\rangle\langle k| \,, \quad \Gamma_{m,n}^\dagger = (X^m Z^n)^\dagger \, Z^{q-n} X^{q-m} = \omega^{mn}\, \Gamma_{-m,-n} \,. \tag{3.19}$$

We also have

$$\text{tr}\,[X^j Z^k X^m Z^n] = q\,\omega^{mk}\,\delta_{j,q-m}\,\delta_{k,q-n} \,, \tag{3.20}$$

where the subscripts on the delta functions are all taken modulo $q$, and this implies for the generic operators, $\Gamma$, the relation

$$\text{tr}\,\left[\Gamma_{j,k}^\dagger \Gamma_{m,n}\right] = \omega^{jk}\,\text{tr}\,\left[\Gamma_{-j,-k}\Gamma_{m,n}\right] = \omega^{mn}\,\text{tr}\,[X^{-j}Z^{-k}X^m Z^n]$$
$$= \omega^{mn}\,\omega^{-mk}\,q\,\delta_{-j,-m}\,\delta_{-k,-n} = q\,\delta_{m,j}\delta_{n,k} \,.$$

which is to say

$$(\,\Gamma\,|\,\Gamma'\,) = \frac{1}{q}\text{tr}\,[\Gamma^\dagger\,\Gamma'] = \delta_{\Gamma,\Gamma'} \,. \tag{3.21}$$

Finally, we can write any matrix, $A$, in the form

$$A = \frac{1}{q}\sum_{j,k=0}^{q-1} \omega^{-jk}\,\text{tr}\,\left[\Gamma_{j,k}^\dagger\,A\right]\Gamma_{j,k} = \sum_{j,k=0}^{q-1} a_{j,k}\,\Gamma_{j,k} \,, \tag{3.22}$$

which follows obviously from completeness of this basis, and the coefficients $a$ have the same inner-product form as Eq. 3.5.

### 3.5. Form of projectors.

The "naïve" projector onto the state $|m\rangle$ is given by

$$P_j^{(m)} = |m\rangle\langle m|_j \equiv \frac{1}{q}\sum_{k=0}^{q-1}\omega^{-mk}\,Z_j^k \,, \tag{3.23}$$

which notably only depends on $Z$, and not $X$, as to be considered proper projectors, they should all share the same basis (and act diagonally therein). The projectors defined in Eq. 3.23 are precisely the "naïve" on-site projectors mentioned in Sec. 3.3.

Another important observation is that we can modify our basis of operators in order to simplify Eq. 3.17. Specifically, of the $q^2$ operators acting on a given site, $q$ of these are given by $\mathbb{1}, Z, \ldots, Z^{q-1}$. We replace these $q$ basis operators (which are simply powers of $Z_j$) by the $q$ different projectors, $P_j^{(m)}$, with appropriate normalization. We define

$$\pi_m = \sqrt{q}\,P^{(m)} = \frac{1}{\sqrt{q}}\sum_{k=0}^{q-1}\omega^{-mk}\,Z_j^k \,, \tag{3.24}$$

which is a linear combination of the $Z$ operators. For $q = 2$, we have $\omega = -1$ and the above simplifies to $(\mathbb{1} \pm Z)/\sqrt{2}$, which are the projectors $u$ $(+)$ and $d$ $(-)$. These projectors satisfy

$$(\,\pi_m\,|\,\pi_n\,) = \frac{1}{q}\text{tr}\,\left[\sqrt{q}P^{(m)}\,\sqrt{q}\,P^{(n)}\right] = \text{tr}\,\left[P^{(m)}\,P^{(n)}\right] = \delta_{m,n} \,, \tag{3.25}$$

9

meaning we normalized correctly.

Note that the other $q^\ell(q^\ell - 1)$ involving at least one power of $X$ on the $\ell$-site cluster need not be modified, and still form a basis for the nondiagonal operators on $q^\ell$ states. The main takeaway from these derivations is that we can write any naïve projector in terms of orthonormal basis operators,

$$P_j^{(m)} = \frac{1}{\sqrt{q}} \pi_{m,j} ~, \tag{3.26}$$

where $\pi_{m,j}$ form a complete, orthonormal subbasis for the set of $q$ operators given by $\left\{ Z_j^k \,\big|\, 0 \leq k < q \right\}$.

### 3.6. Diagonal versus nondiagonal operators

We now revisit the second term in Eq. 3.17, corresponding to single-state blocks of Eq. 1.1 that are not associated with a Haar random unitary (as they have no dynamics). We have

$$\left( \sigma_r^\mu \,\big|\, T_r^{(2)} \,\big|\, \sigma_r^\nu \right) = \frac{1}{q^\ell} \sum_{\beta,\beta'} \operatorname{tr}\left[ \sigma_r^\mu \, P_r^\beta \, \sigma_r^\nu \, P_r^{\beta'} \right] \tag{3.27}$$

$$= \frac{1}{q^\ell} \sum_\beta \operatorname{tr}\left[ \sigma_r^\mu \, P_r^\beta \, \sigma_r^\nu \, P_r^\beta \right] + \frac{1}{q^\ell} \sum_{\beta \neq \beta'} \operatorname{tr}\left[ \sigma_r^\mu \, P_r^\beta \, \sigma_r^\nu \, P_r^{\beta'} \right] \tag{3.28}$$

and note that each projector in this term is "naïve" by construction, meaning ($i$) we can write $P_r^\beta = |\beta_r\rangle\langle\beta_r|$ and ($ii$) we can insert a factor of $1/n_\beta = 1$ in the first summand. This gives

$$\left( \sigma_r^\mu \,\big|\, T_r^{(2)} \,\big|\, \sigma_r^\nu \right) = \frac{1}{q^\ell} \sum_\beta \frac{1}{n_\beta} \operatorname{tr}\left[ \sigma_r^\mu \, |\beta_r\rangle\langle\beta_r| \, \sigma_r^\nu \, P_r^\beta \, |\beta_r\rangle\langle\beta_r| \right] + \frac{1}{q^\ell} \sum_{\beta \neq \beta'} \operatorname{tr}\left[ \sigma_r^\mu \, |\beta_r\rangle\langle\beta_r| \, \sigma_r^\nu \, |\beta_r'\rangle\langle\beta_r'| \right]$$

$$= \frac{1}{q^\ell} \sum_\beta \frac{1}{n_\beta} \, \langle\beta_r|\sigma_r^\mu|\beta_r\rangle \, \langle\beta_r|\sigma_r^\nu|\beta_r\rangle + \frac{1}{q^\ell} \sum_{\beta \neq \beta'} \langle\beta_r'|\sigma_r^\mu|\beta_r\rangle \, \langle\beta_r|\sigma_r^\nu|\beta_r'\rangle ~, \tag{3.29}$$

where it is clear from the form above that the first term is nonzero only if both $\sigma_r^\mu$ and $\sigma_r^\nu$ are *diagonal* operators (i.e., in the operator subspace spanned by $Z_r^m$, or equivalently, the projectors $\pi_{m,r}$). The second term, however, is nonzero only if $\sigma_r^\mu$ and $\sigma_r^\nu$ are *both* nondiagonal operators. The fact that both operators in each term must be part of the same subset implies that there is no mixing between these blocks: The transfer matrix gates, $T_r$, map diagonal operators to other diagonal operators, and nondiagonal operators to other nondiagonal operators. We now rewrite the above as

$$\left( \sigma_r^\mu \,\big|\, T_r^{(2)} \,\big|\, \sigma_r^\nu \right) = \frac{1}{q^\ell} \sum_\beta \frac{1}{n_\beta} \operatorname{tr}\left[ P_r^\beta \, \sigma^\mu \right] \operatorname{tr}\left[ P_r^\beta \, \sigma^\nu \right] + \frac{1}{q^\ell} \sum_{\beta \neq \beta'} \operatorname{tr}\left[ \sigma_r^\mu \, P_r^\beta \, \sigma_r^\nu \, P_r^{\beta'} \right] ~, \tag{3.30}$$

and note that the first term above is identical to the terms in Eq. 3.17 that resulted from blocks, $\alpha$, associated with Haar unitaries:

$$\left( \sigma_r^\mu \,\big|\, T_r^{(1)} \,\big|\, \sigma_r^\nu \right) = \tfrac{1}{q^\ell} \sum_\alpha \tfrac{1}{n_\alpha} \operatorname{tr}\left[ P_r^\alpha \, \sigma^\mu \right] \operatorname{tr}\left[ P_r^\alpha \, \sigma^\nu \right] ~, \tag{3.31}$$

where $\alpha$ corresponds to blocks associated with nontrivial dynamics (and therefore Haar unitaries), while $\beta$ corresponds to trivial blocks containing a single, dynamically trivial state. The full expression for a transfer matrix gate acting on cluster $r$ is given by $T_r = T_r^{(1)} + T_r^{(2)}$.

### 3.7. General result and connection to the spectral form factor's transfer matrix

Here we consider the form of the transfer matrix gates, $T_r$. The full transfer matrix, $\mathcal{T}$, for a single time step is a circuit with identical structure as $\mathcal{W}_t$ (the circuit that evolves the system by one time step) with the unitary gates $\mathcal{U}_r$ replaced by Hermitian gates $T_r$. Returning to Eq. 3.17, it is convenient to restrict to one of the two operator sectors (diagonal versus nondiagonal operators).

*a. Nondiagonal operators.* Let us first consider operators that contain one or more powers of the operator $X$. These operators are nondiagonal, which means that the only terms that contribute to $T$ are the second terms in Eq. 3.30, i.e.

$$(\sigma_r^\mu | T_r | \sigma_r^\nu) = \frac{1}{q^\ell} \sum_{\beta \neq \beta'} \text{tr}\left[\sigma_r^\mu P_r^\beta \sigma_r^\nu P_r^{\beta'}\right] = \frac{1}{q^{2\ell}} \sum_{\beta \neq \beta'} \text{tr}\left[\sigma_r^\mu \pi_r^\beta \sigma_r^\nu \pi_r^{\beta'}\right] , \quad (3.32)$$

and clearly, nondiagonal operators always map to other nondiagonal operators. Also recall that $\beta$ only includes trivial blocks (with no associated unitary dynamics), each having only one state. Further simplification of Eq. 3.32 does not appear to be possible without loss of generality, and the general form is

$$T_r = \sum_{\mu_r, \nu_r} \frac{1}{q^{2\ell}} \sum_{\beta \neq \beta'} \text{tr}\left[\sigma_r^{\mu_r} \pi_r^\beta \sigma_r^{\nu_r} \pi_r^{\beta'}\right] |\sigma_r^{\mu_r})(\sigma_r^{\nu_r}| . \quad (3.33)$$

*b. Diagonal operators.* For diagonal operators, which contain only the identity and powers of $Z$, both $T_r^{(1)}$ and the first term in $T_r^{(2)}$ contribute:

$$(\sigma_r^\mu | T_r | \sigma_r^\nu) = \frac{1}{q^\ell} \sum_\alpha \frac{1}{n_\alpha} \text{tr}\left[P_r^\alpha \sigma^\mu\right] \text{tr}\left[P_r^\alpha \sigma^\nu\right] + \frac{1}{q^\ell} \sum_\beta \frac{1}{n_\beta} \text{tr}\left[P_r^\beta \sigma^\mu\right] \text{tr}\left[P_r^\beta \sigma^\nu\right] , \quad (3.34)$$

which we can combine into a single sum over *all* distinct blocks, labelled by $b$:

$$(\sigma_r^\mu | T_r | \sigma_r^\nu) = \frac{1}{q^\ell} \sum_b \frac{1}{n_b} \text{tr}\left[P_r^b \sigma^\mu\right] \text{tr}\left[P_r^b \sigma^\nu\right] \quad (3.35)$$

$$= \sum_b \frac{1}{n_b} \sum_{m, m' \in b} (\sigma_r^\mu | \pi_r^m)(\pi_r^{m'} | \sigma_r^\nu) ,$$

and finally, we have

$$T_r = \sum_b \frac{1}{n_b} \sum_{m, m' \in b} |\pi_r^m)(\pi_r^{m'}| , \quad (3.36)$$

which has the same form as the gates that appear in the transfer matrix corresponding to the spectral form factor (2.4), except instead of acting on $q$-state spins, it acts on the subspace of operators comprising the projective basis, $\{\pi_r^m\}$, of which there are $q$ per site. Each block, $b$, has uniform entries with value $1/n_b$, where $n_b$ is the size of the block, and includes all blocks (both trivial and nontrivial dynamics).

In other words, each of the $n \times n$ blocks is a square matrix with unit trace and all matrix elements the same, nonzero value $(1/n)$. For $n > 1$, every projector that is part of the block gets mapped to an equal superposition of all $n$ projectors, each with weight $1/n$, upon applying the transfer matrix. Hence, the Thouless time extracted from the spectral form factor calculation is the same as the one that recovers from correlation functions in all circuit models. In both cases, this time is the inverse of the gap of a transfer matrix (or equivalently, the corresponding "Hamiltonian"), and reflects the amount of time it takes for information to spread throughout the entire system.

Such a relation between the spectral form factor—itself a two-point correlator—and generic correlation functions of local observables was first suggested in Ref. 12. Here, we provide a rigorous derivation of this result; while the exact equivalence between these two quantities is likely an artifact of the chaotic circuits used in this derivation, we expect that the relation itself between the two quantities—and the equivalence of their characteristic "Thouless times"—is a generic feature of chaotic, many-body quantum systems.

### 3.8. Transfer matrices for particular cases

Starting from Eq. 3.10 we have

$$\mathcal{T}_{\mu,\nu}^{(\lambda)} = \left(\sigma^\mu \left| \mathcal{T}_{(\lambda)} \right| \sigma^\nu\right) = \frac{1}{\mathcal{D}} \text{tr}\left[\sigma^\mu \overline{\mathcal{W}_\lambda^\dagger \sigma^\nu \mathcal{W}_\lambda}\right] , \quad (3.37)$$

11

where, for notational convenience, the $\lambda$ subscript labels different layers of the circuit corresponding to a single "time step".

For models with symmetries and/or constraints encoded via projectors, the form of the transfer matrix is given by Eq. 3.36 in the subspace of diagonal operators, and by Eq. 3.33 for nondiagonal operators.

*a. Basic model.* For a standard brickwork circuit with no conservation laws or constraints, a single time step consists of two layers, corresponding to even versus odd bonds. Taking the local Hilbert space to have dimension $q$, so the many-body Hilbert space dimension is $\mathcal{D} = q^L$, the corresponding transfer matrices (3.37) are given by

$$\mathcal{T}^{e/o}_{\mu,\nu} = \frac{1}{q^L} \operatorname{tr}\left[\sigma^\mu \,\overline{\mathcal{W}^\dagger_{e/o} \sigma^\nu \mathcal{W}_{e/o}}\,\right] \tag{3.38a}$$

$$= \prod_{j \in e/o} \frac{1}{q^2} \operatorname{tr}\left[\sigma^{\mu_j}_j \sigma^{\mu_{j+1}}_{j+1} \,\overline{U^\dagger_{j,j+1} \sigma^{\nu_j}_j \sigma^{\nu_{j+1}}_{j+1} U_{j,j+1}}\,\right] \tag{3.38b}$$

where $U$ represents a $q^2 \times q^2$ Haar unitary, and we used the result for averaging a onefold Haar channel over the Circular Unitary Ensemble [1]. Now,

$$\mathcal{T}^{e/o}_{\mu,\nu} = \prod_{j \in e/o} \frac{1}{q^4} \operatorname{tr}\left[\sigma^{\mu_j}_j \sigma^{\mu_{j+1}}_{j+1}\right] \operatorname{tr}\left[\sigma^{\nu_j}_j \sigma^{\nu_{j+1}}_{j+1}\right] \tag{3.38c}$$

$$= \prod_j \frac{1}{q^2} \operatorname{tr}\left[\sigma^{\mu_j}_j\right] \operatorname{tr}\left[\sigma^{\nu_j}_j\right] = \prod_j \delta_{\mu_j,0}\,\delta_{\nu_j,0} \quad, \tag{3.38d}$$

which implies

$$\mathcal{T}_{e/o} = |\mathbb{1})(\mathbb{1}| \quad, \tag{3.38e}$$

which leaves the [many-body] identity operator unchanged, and annihilates *any* other operators. Thus, a fully random Haar unitary circuit evolution instantly kills *all* nontrivial correlations after a single time step: The timescale on which correlations decay is order one, just as the "spectral" Thouless time is one (the spectral form factor shows a linear ramp in this model after a single circuit layer step, i.e. $t_{\text{Th}} = 1$) [5].

*b. $U(1)$ conserving case.* Here we again have two-site unitary gates comprising our circuit, with layers acting on even/odd (e/o) bonds. Each site contains a $q$-dimensional spin, meant to facilitate Haar averaging, and a spin $1/2$ (qubit), which encodes the $U(1)$ conservation law. Particularly, every gate conserves the total $z$ component of spin, $S^z_{j,j+1} = \sigma^z_j + \sigma^z_{j+1}$.

Ignoring the $q$-dit operator content (which is simply $|\mathbb{1})(\mathbb{1}|$), each Hermitian gate in $\mathcal{T}$ takes the form

$$\begin{pmatrix} 1 & 0 & 0 & 0 \\ 0 & \frac{1}{2} & \frac{1}{2} & 0 \\ 0 & \frac{1}{2} & \frac{1}{2} & 0 \\ 0 & 0 & 0 & 1 \end{pmatrix} \quad \text{acting on} \quad \begin{pmatrix} u_j\,u_{j+1} \\ u_j\,d_{j+1} \\ d_j\,u_{j+1} \\ d_j\,d_{j+1} \end{pmatrix} \quad, \tag{3.39}$$

where the connection to the transfer matrix that appears in the calculation of the spectral form factor is apparent upon taking $|u) \to |\bullet\rangle$ and $|d) \to |\circ\rangle$ [9].

We note that the above transfer matrix gate, $T_{j,j+1}$ can be written as $T_{j,j+1} = \mathbb{1}_{j,j+1} - H_{j,j+1}$ where, in the spin-$1/2$ language, $H_{j,j+1}$ is the $SU(2)$ symmetric XXX Heisenberg Hamiltonian,

$$H_{j,j+1} = -\frac{1}{4}\left(\vec{\sigma}_j \cdot \vec{\sigma}_{j+1} - \mathbb{1}\right) \quad,$$

consistent with a gap of $k^2 \propto L^{-2}$, and Thouless time $t_{\text{Th}} \propto L^2$, indicating diffusion. Additionally, we note that the brickwork transfer matrix, $\mathcal{T}$, is also integrable, with solutions given by the Algebraic Bethe Ansatz (see Supplemental Material for Ref. 9 for more details on the ABA construction).

c. *Fredkin circuit.* Here we go through the procedure for the Fredkin circuit presented in the main text. The Fredkin model is also defined on a chain of qubits. Hopping can occur between sites $j$ and $j+1$ if site $j-1$ is unoccupied or hopping can occur between sites $j-1$ and $j$ if site $j+1$ is occupied (see Fig. 2). The unitary describing this evolution on a given set of bonds can be written as $\mathcal{F} = \mathcal{F}_L \mathcal{F}_R$ where $\mathcal{F}_{L/R}$ are each a product over $L/R$ gates. For cluster $r = \{j-1, j, j+1\}$, the gates are

$$\mathcal{U}_{r,L} = |\bullet\rangle\langle\bullet|_{j-1} \mathcal{U}^{(L)}_{j,j+1} + |\circ\rangle\langle\circ|_{j-1} \tag{3.40}$$

$$\mathcal{U}_{r,R} = \mathcal{U}^{(R)}_{j-1,j} |\circ\rangle\langle\circ|_{j+1} + |\bullet\rangle\langle\bullet|_{j+1} \ , \tag{3.41}$$

where $\mathcal{U}^{(R)}_{j-1,j}$ and $\mathcal{U}^{(L)}_{j,j+1}$ are two-site unitary gates that preserves the $U(1)$ particle number, and each charge sector of the unitary is an independently drawn random Haar unitary (see Fig. 1).

Following the procedure of Sec. 3, we construct the transfer matrix gates in the basis of diagonal operators (i.e., projectors onto the $\bullet$ and $\circ$ states),

$$T_r^{(L)} = \begin{bmatrix} 1 & 0 \\ 0 & 0 \end{bmatrix} \otimes \begin{bmatrix} 1 & 0 & 0 & 0 \\ 0 & 1/2 & 1/2 & 0 \\ 0 & 1/2 & 1/2 & 0 \\ 0 & 0 & 0 & 1 \end{bmatrix} + \begin{bmatrix} 0 & 0 \\ 0 & 1 \end{bmatrix} \otimes \mathbb{1}_{j,j+1} \tag{3.42}$$

$$T_r^{(R)} = \begin{bmatrix} 1 & 0 & 0 & 0 \\ 0 & 1/2 & 1/2 & 0 \\ 0 & 1/2 & 1/2 & 0 \\ 0 & 0 & 0 & 1 \end{bmatrix} \otimes \begin{bmatrix} 0 & 0 \\ 0 & 1 \end{bmatrix} + \mathbb{1}_{j-1,j} \otimes \begin{bmatrix} 1 & 0 \\ 0 & 0 \end{bmatrix} \ , \tag{3.43}$$

Note that the dimension of the transfer matrices is $2^3 \times 2^3$ since there are only two basis elements per site that are charge zero (i.e., diagonal). Lastly, because these two gates commute with each other, we can combine them into a single gate, $T_r = T_r^{(L)} T_r^{(R)}$.

d. *"U(1) East" circuit.* The $U(1)$ East model also acts on a chain of qubits, and allows hopping on sites $j-1$ and $j$ if site $j+1$ is occupied ($\bullet$). Thus the unitary describing the evolution is simply the "$R$" (or "East") gate from the Fredkin model (3.41) and the transfer matrix is given by Eq. 3.43, i.e. for $r = \{j-1, j, j+1\}$ we have

$$T_r = \begin{bmatrix} 1 & 0 & 0 & 0 \\ 0 & 1/2 & 1/2 & 0 \\ 0 & 1/2 & 1/2 & 0 \\ 0 & 0 & 0 & 1 \end{bmatrix} \otimes \begin{bmatrix} 0 & 0 \\ 0 & 1 \end{bmatrix} + \mathbb{1}_{j-1,j} \otimes \begin{bmatrix} 1 & 0 \\ 0 & 0 \end{bmatrix} \ . \tag{3.44}$$

e. *"GLT" circuit.* The Gonçalves-Landim-Toninelli (GLT) model [16] also acts on a chain of qubits, and allows hopping between sites $j$ and $j+1$ if *either* neighbor ($j-1$ or $j+2$) is occupied ($\bullet$). Like the Fredkin model, we use separate unitary gates, labelled $L$ and $R$ to denote the constraint coming from the left ($L$) or right ($R$) neighbor, i.e. which for cluster $r = \{j-1, j, j+1\}$ are given by

$$\mathcal{U}_{r,L} = |\bullet\rangle\langle\bullet|_{j-1} \mathcal{U}^{(L)}_{j,j+1} + |\circ\rangle\langle\circ|_{j-1} \tag{3.45}$$

$$\mathcal{U}_{r,R} = \mathcal{U}^{(R)}_{j-1,j} |\bullet\rangle\langle\bullet|_{j+1} + |\circ\rangle\langle\circ|_{j+1} \ , \tag{3.46}$$

where, as for Fredkin, $\mathcal{U}^{(R)}_{j-1,j}$ and $\mathcal{U}^{(L)}_{j,j+1}$ are two-site $U(1)$ symmetric gates, whose blocks are independently drawn Haar unitaries (per Fig. 1). The transfer matrix gates are given by

$$T_r^{(L)} = \begin{bmatrix} 1 & 0 \\ 0 & 0 \end{bmatrix} \otimes \begin{bmatrix} 1 & 0 & 0 & 0 \\ 0 & 1/2 & 1/2 & 0 \\ 0 & 1/2 & 1/2 & 0 \\ 0 & 0 & 0 & 1 \end{bmatrix} + \begin{bmatrix} 0 & 0 \\ 0 & 1 \end{bmatrix} \otimes \mathbb{1}_{j,j+1} \tag{3.47}$$

$$T_r^{(R)} = \begin{bmatrix} 1 & 0 & 0 & 0 \\ 0 & 1/2 & 1/2 & 0 \\ 0 & 1/2 & 1/2 & 0 \\ 0 & 0 & 0 & 1 \end{bmatrix} \otimes \begin{bmatrix} 1 & 0 \\ 0 & 0 \end{bmatrix} + \mathbb{1}_{j-1,j} \otimes \begin{bmatrix} 0 & 0 \\ 0 & 1 \end{bmatrix} \ . \tag{3.48}$$

*f.* *"PXYP" circuit.* The PXYP model acts on a chain of qubits, with hopping between sites $j$ and $j+1$ allowed only if *both* neighbors ($j-1$ and $j+2$) are occupied ($\bullet$). The corresponding unitary gate acts on four sites as

$$\mathcal{U}_r = |\bullet\rangle\langle\bullet|_{j-1}\mathcal{U}_{j,j+1}|\bullet\rangle\langle\bullet|_{j+2} + |\circ\rangle\langle\circ|_{j-1}|\bullet\rangle\langle\bullet|_{j+2} + |\bullet\rangle\langle\bullet|_{j-1}|\circ\rangle\langle\circ|_{j+2} + |\circ\rangle\langle\circ|_{j-1}|\circ\rangle\langle\circ|_{j+2} \;, \quad (3.49)$$

where $\mathcal{U}_{j,j+1}$ is the two-site $U(1)$ gate that appears in Fig. 1. The corresponding transfer matrix for diagonal operators is given by

$$T_r = \begin{bmatrix} 1 & 0 \\ 0 & 0 \end{bmatrix} \otimes \begin{bmatrix} 1 & 0 & 0 & 0 \\ 0 & 1/2 & 1/2 & 0 \\ 0 & 1/2 & 1/2 & 0 \\ 0 & 0 & 0 & 1 \end{bmatrix} \otimes \begin{bmatrix} 1 & 0 \\ 0 & 0 \end{bmatrix}$$
$$+ \begin{bmatrix} 1 & 0 \\ 0 & 0 \end{bmatrix} \otimes \mathbb{1}_{j,j+1} \otimes \begin{bmatrix} 0 & 0 \\ 0 & 1 \end{bmatrix} + \begin{bmatrix} 0 & 0 \\ 0 & 1 \end{bmatrix} \otimes \mathbb{1}_{j,j+1} \otimes \begin{bmatrix} 1 & 0 \\ 0 & 0 \end{bmatrix} + \begin{bmatrix} 0 & 0 \\ 0 & 1 \end{bmatrix} \otimes \mathbb{1}_{j,j+1} \otimes \begin{bmatrix} 0 & 0 \\ 0 & 1 \end{bmatrix} \;. \quad (3.50)$$

*g.* *"XNOR" circuit.* The XNOR model acts on a chain of qubits, with hopping between sites $j$ and $j+1$ allowed only if *both* neighbors ($j-1$ and $j+2$) are occupied ($\bullet$) or unoccupied ($\circ$). The corresponding unitary gate acts on four sites as

$$\mathcal{U}_r = |\bullet\rangle\langle\bullet|_{j-1}\mathcal{U}^\bullet_{j,j+1} + |\circ\rangle\langle\circ|_{j-1}\mathcal{U}^\circ_{j,j+1}|\circ\rangle\langle\circ|_{j+2}|\bullet\rangle\langle\bullet|_{j+2} + |\circ\rangle\langle\circ|_{j-1}|\bullet\rangle\langle\bullet|_{j+2} + |\bullet\rangle\langle\bullet|_{j-1}|\circ\rangle\langle\circ|_{j+2} \;, \quad (3.51)$$

where $\mathcal{U}^\bullet_{j,j+1}$ and $\mathcal{U}^\circ_{j,j+1}$ are the two-site $U(1)$ gates that appears in Fig. 1. The corresponding transfer matrix for diagonal operators is given by

$$T_r = \begin{bmatrix} 1 & 0 \\ 0 & 0 \end{bmatrix} \otimes \begin{bmatrix} 1 & 0 & 0 & 0 \\ 0 & 1/2 & 1/2 & 0 \\ 0 & 1/2 & 1/2 & 0 \\ 0 & 0 & 0 & 1 \end{bmatrix} \otimes \begin{bmatrix} 1 & 0 \\ 0 & 0 \end{bmatrix} + \begin{bmatrix} 0 & 0 \\ 0 & 1 \end{bmatrix} \otimes \begin{bmatrix} 1 & 0 & 0 & 0 \\ 0 & 1/2 & 1/2 & 0 \\ 0 & 1/2 & 1/2 & 0 \\ 0 & 0 & 0 & 1 \end{bmatrix} \otimes \begin{bmatrix} 0 & 0 \\ 0 & 1 \end{bmatrix}$$
$$+ \begin{bmatrix} 1 & 0 \\ 0 & 0 \end{bmatrix} \otimes \mathbb{1}_{j,j+1} \otimes \begin{bmatrix} 0 & 0 \\ 0 & 1 \end{bmatrix} + \begin{bmatrix} 0 & 0 \\ 0 & 1 \end{bmatrix} \otimes \mathbb{1}_{j,j+1} \otimes \begin{bmatrix} 1 & 0 \\ 0 & 0 \end{bmatrix} \;. \quad (3.52)$$

*h.* *"Motzkin" circuit.* The Motzkin model acts on a chain of spins one, with dynamics generated by a two-site unitary whose only constraint is that no state can transition to the Motzkin path, $\searrow\!\!\nearrow$, where the correspondence between the pictorial path notation and the standard spin one notation is given by $\{-1, 0, +1\} = \{\searrow, \_, \nearrow\}$. The corresponding unitary is given by

$$\mathcal{U}_{j,j+1} = \sum_{Q=\pm 1, \pm 2} P^{(Q)}_{j,j+1} U^{(Q)}_{j,j+1} P^{(Q)}_{j,j+1}$$
$$+ (\mathbb{1}_{j,j+1} - |\searrow\!\!\nearrow\rangle\langle\searrow\!\!\nearrow|) P^{(0)}_{j,j+1} U^{(0)}_{j,j+1} P^{(0)}_{j,j+1} (\mathbb{1}_{j,j+1} - |\searrow\!\!\nearrow\rangle\langle\searrow\!\!\nearrow|) + |\searrow\!\!\nearrow\rangle\langle\searrow\!\!\nearrow|, \quad (3.53)$$

where $P^{(Q)}_{j,j+1}$ projects onto states of sites $j$ and $j+1$ with charge $Q = n_j + n_{j+1}$, and $U^{(Q)}_{j,j+1}$ are independently drawn Haar unitaries whose dimensions are equal to the subspace size, except for $U^{(0)}_{j,j+1}$, which is a $2 \times 2$ Haar unitary due to the constraint. In general, the size of $U$ is given by the number of states not annihilated by the combination of projectors sandwiching it.

The corresponding transfer matrix is given by,

$$T_{j,j+1} = \frac{1}{2} \sum_{Q=\pm 1, \pm 2} |Q| \, P^{(Q)}_{j,j+1}$$
$$+ \frac{1}{2}(\mathbb{1}_{j,j+1} - |\searrow\!\!\nearrow\rangle\langle\searrow\!\!\nearrow|) P^{(0)}_{j,j+1}(\mathbb{1}_{j,j+1} - |\searrow\!\!\nearrow\rangle\langle\searrow\!\!\nearrow|) + |\searrow\!\!\nearrow\rangle\langle\searrow\!\!\nearrow|, \quad (3.54)$$

The Motzkin model is closely related to the Fredkin model. We show in Section. 5.3 that the dynamics of the Motzkin circuit lie in the same universality class as the Fredkin model, exhibiting $z \simeq 8/3$ subdiffusion.

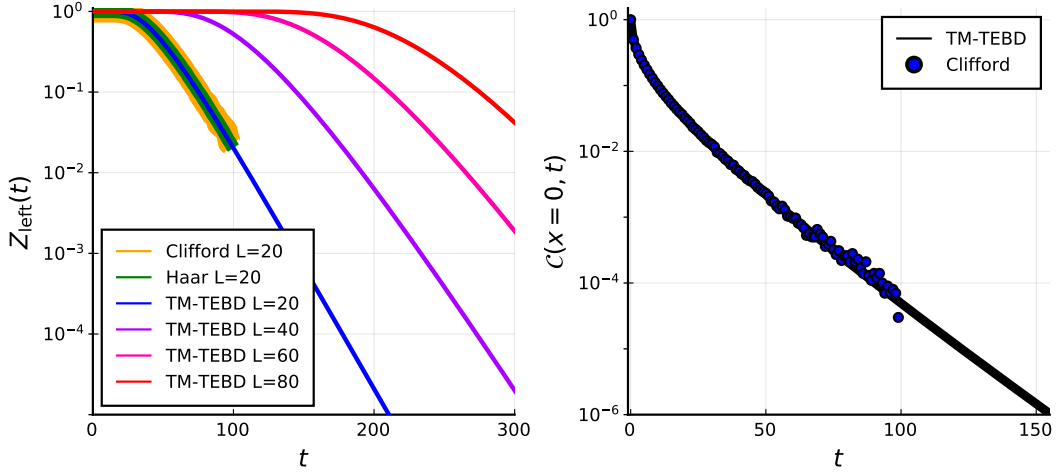

FIG. 4. **East circuit.** *Left*: Expectation value of $Z_{\text{left}}(t)$ of the leftmost site in the East circuit, starting from the state where only the rightmost state is occupied at $t=0$. *Right*: The Haar averaged infinite temperature correlation function, $\mathcal{C}(x,t) = \mathcal{D}^{-1} \operatorname{tr}\left[\overline{Z_0(t) Z_x(0)}\right]$ using TEBD on the Haar averaged transfer matrix and Clifford numerics.

## 4. NONCONSERVING CASE: THE EAST CIRCUIT

The East circuit (not to be confused with the "$U(1)$ East" model) also acts on a chain of $L$ qubits, using two-site gates that evolve site $j$ only if site $j+1$ (the "East" neighbor) is occupied ($\bullet$). These gates take the form

$$\mathcal{U}_r = U_j \, |\bullet\rangle\langle\bullet|_{j+1} + |\circ\rangle\langle\circ|_{j+1} \quad , \tag{4.1}$$

where $U$ is a random $2 \times 2$ Haar unitary acting on site $j$ alone.

This model has *no local conservation laws*. However, states where site $x$ is occupied and all sites to the right of $x$ are unoccupied will continue to have $x$ be the rightmost occupied site for all time. One expects the states on sites $1, \ldots, x$ to thermalize (i.e., reach the steady state, a uniform superposition of occupied / unoccupied) in $O(1)$ time, and so the thermalization time of these states is essentially unaffected by the constraint in the region outside the sea of down spins.

Nonetheless, there are states whose thermalization is delayed under the evolution of Eq. 4.1. These states have all down spins for sites $1, \ldots, x$ (with $x < L$), followed by all up spins. The time it takes for the entire region of down spins to be affected by the unitary gates is $O(x)$. However, once the unitary gates act on a given set of spins that satisfy the constraint, the time it takes to thermalize is independent of system size. This delay to thermalization is similar to the physics of local quenches, and the $O(x)$ timescale should not be associated with the system slowly thermalizing, but rather a "light cone delay". We confirm that the time to thermalize is independent of system size by computing the gap of the transfer matrix whose gates are given by,

$$T_{j,j+1} = \frac{1}{2} \begin{bmatrix} 1 & 1 \\ 1 & 1 \end{bmatrix} \otimes \begin{bmatrix} 1 & 0 \\ 0 & 0 \end{bmatrix} + \begin{bmatrix} 1 & 0 \\ 0 & 1 \end{bmatrix} \otimes \begin{bmatrix} 0 & 0 \\ 0 & 1 \end{bmatrix} \quad . \tag{4.2}$$

We also compute correlation functions, $\mathcal{C}(x,t) \equiv \mathcal{D}^{-1} \operatorname{tr}\left[\overline{Z_0(t) Z_x}\right]$ and extract the same information. Note that in this case we can replace Haar averages by averages over Pauli matrices, and the unitary gates (4.1) are a subset of Clifford gates, which map Pauli strings to Pauli strings. Thus, we only need to evolve Pauli strings to simulate the circuit. In addition, the time-evolved form of $Z_0(t)$ will always be a Pauli string that acts on site zero as $Z$, and thus $\operatorname{tr}\left[\overline{Z_0(t) Z_x}\right]$ is guaranteed to vanish *unless* $x=0$. Simulating the

15

action of Clifford gates on Pauli strings can be implemented efficiently; however, at later times we find that more averaging is required to obtain sufficiently smooth data, and thus using TEBD to simulate the transfer matrix is still extremely useful.

Our results are shown in Fig. 4. In the first panel we examine $\overline{\langle\psi_{\text{slow}}|Z_{\text{left}}(t)|\psi_{\text{slow}}\rangle}$, where $|\psi_{\text{slow}}\rangle$ is the state where the rightmost site is occupied (●) with all other sites unoccupied (○). First, we show that we get identical results evolving by the unitary gates given in Eq. 4.1 and then Haar averaging (using TEBD), using the Clifford evolution of Pauli strings, and directly simulating the transfer matrix (4.2) using TEBD. Second, in the first panel we see the aforementioned $O(L)$ "light cone delay", followed by an exponential decay with finite decay time. In the second panel, we examine the infinite temperature correlation functions, $C(x,t)$, which again show an exponential decay, but do not feature the $O(x)$ delay time since $\left\langle\overline{Z_{\text{left}}(t)}\right\rangle$ has exponentially small overlap with such states.

## 5. ADDITIONAL RESULTS

### 5.1. $U(1)$ East and PXYP constraints

The slow dynamics in the $U(1)$ East and PXYP circuit models is due to Hilbert space *fragmentation* — that is, the dynamics decouples into exponentially many sectors in a way that effectly bottlenecks the spread of local quantities across the system. Previously studied fragmented systems have been classified into *strongly* fragmented and *weakly* fragmented subclasses [11, 17, 18]. In weakly fragmented systems, typical eigenstates thermalize, whereas in strongly fragmented systems they do not. Based on their studies of a weakly fragmented system, Ref. 17 proposed using the existence of large sectors that fill a large fraction of Hilbert space as a diagnostic for weak fragmentation. The logic behind this diagnostic is that within the large sector, the dynamics does not suffer bottlenecks and thermalizes normally. While other sectors may not experience thermalization and instead exhibit *quantum scar* behavior [19], thermalization in the large sector is sufficient for thermalization *on average* for typical states [17].

We find that the same diagnostic correctly identifies strong and weak fragmentation in the PXYP and $U(1)$ East models, respectively. Using brute force counting for small system sizes, $L \leq 20$, we compute the number of sectors and the size of the largest sector for both models. Using Ref. 20, we found integer sequences matching each of these computations and more efficient methods for continuing the sequence. The results are summarized in Fig. 6. For a Hilbert space of size $2^L$ and with $\mathcal{N}$ sectors, the average size $\langle N\rangle$ of the sectors is $\langle N\rangle = 2^L/\mathcal{N}$. The plot shows $\langle N\rangle/2^L = 1/\mathcal{N}$. For both the PXYP and the $U(1)$ East model, the number of sectors grows exponentially $\mathcal{N} \sim b^L$ with $1 < b < 2$. (The precise values for $b$ are reported below.) The number of sectors in the Fredkin model $\mathcal{N} \sim L^2$ is plotted for comparison. In contrast, the behavior of the size $N$ of the largest sector distinguishes the PXYP and $U(1)$ East models. For the PXYP model, the size of the largest sector, $N$, grows exponentially $N \sim c^L$ with a base $c < 2$, while the $U(1)$ East model has a largest sector of size $N \sim 2^L/L$. This sector thus consists of a polynomially decaying fraction of the corresponding $U(1)$ charge sector rather than an exponentially decaying fraction. These observations along with that of thermalization exhibited in the average correlation functions indicate that the $U(1)$ East model is weakly fragmented, while the PXYP model is strongly fragmented. In particular, for $U(1)$ East the charge profiles continue to broaden over time (shown in Fig. 6) and the return probability decays over time (Fig. 5) with no saturation. Intriguingly, the form of this decay appears to be slower than any power law but faster than logarithmic. In contrast, the profiles for the PXYP model (shown in Fig. 6) are localized, with the variance (main text) and return probability (Fig. 5) saturating quickly in time.

The strong fragmentation of the PXYP model can be anticipated from the form of the constraints. For example, the presence of two or more consecutive holes is completely frozen in the dynamics, with no particles allowed to hop into or across that stretch of holes for all time. This intuitively accounts for the saturation of the return probability shown in Fig. 5, as a finite fraction of configurations will never experience hopping into or from a fixed site $x$. The number of decoupled sectors, given by the number of eigenvalues one in the transfer matrix (3.50), grows exponentially with system size.

Our numerical investigation into the sectors of the PXYP transfer matrix reveals that each sector can be labeled by a unique representative state in which no particles can hop to the right. Such a state has no four

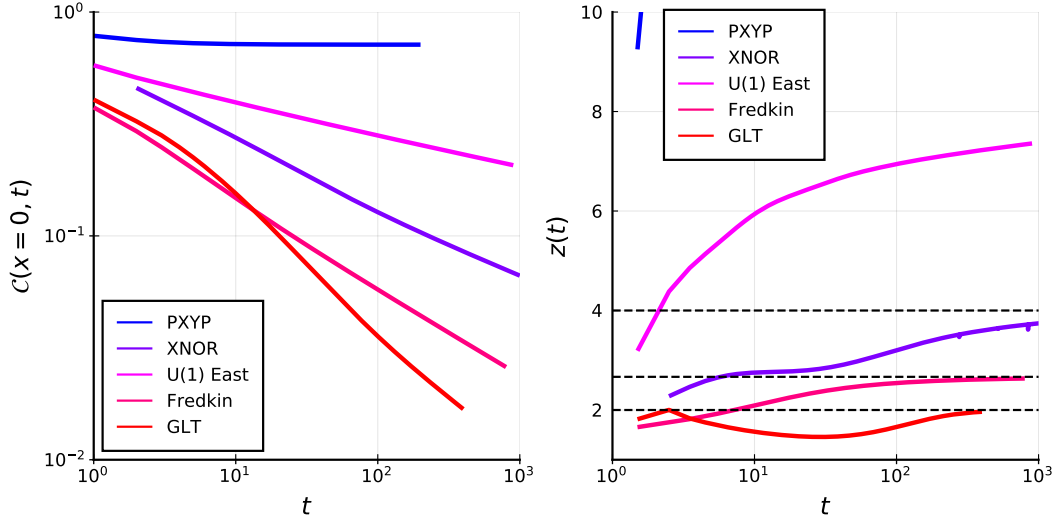

FIG. 5. **Return Probabilities.** *Left:* Return probabilities $\mathcal{C}(x=0,t) \sim t^{-1/z}$ for the PXYP, $U(1)$ East, XNOR, Fredkin and GLT models, computed via TEBD of the transfer matrix. *Right:* Putative time-dependent dynamical exponent, $z(t)$, extracted from the return probability as $-1/z(t) = d \log \mathcal{C}(0,t)/d \log t$. For GLT, Fredkin, and XNOR, we find values compatible with those obtained from the variance of the profiles (main text) of $\mathcal{C}(x,t)$: $z = 8/3$, $z = 2$, and $z = 4$, respectively (dashed lines). For PXYP and $U(1)$ East, the value of $z(t)$ is different from that extracted from the variance, and continues to increase with time, $t$. As the return probability saturates for PXYP but not for the $U(1)$ East model, they are, respectively, well described as localized and quasilocalized.

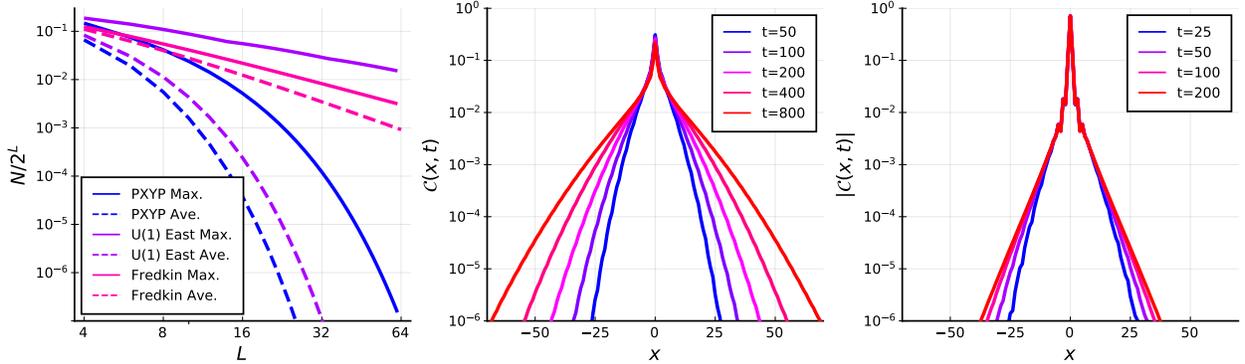

FIG. 6. **Strong vs. weak fragmentation.** *Left:* Average and maximum sector size (normalized by $2^L$) for three constrained models. For PXYP, both the average and maximum sector size scale exponentially slower than $2^L$, signaling *strong* fragmentation. For $U(1)$ East, the average sector size scales exponentially slower than $2^L$, while the growth of the maximum sector size with $L$ is only *polynomially* slower than $2^L$, signaling *weak* fragmentation. For comparison, the average and maximum sector sizes of the unfragmented Fredkin model grow polynomially slower than $2^L$. *Middle:* Correlation profiles for the $U(1)$ East model; the width continues to grow without bound (although slowly), and the return probability slowly decays. *Right:* Correlation profiles for the PXYP model; the width and the return probability saturate with time.

consecutive sites with the configuration ●●○●. This number of sectors can be identified as the sequence A049864 in Ref. 20, which reports a recursion relation for the number of sectors $\mathcal{N}_L = 2\mathcal{N}_{L-1} - \mathcal{N}_{L-3} + \mathcal{N}_{L-4}$, with $\mathcal{N}_1 = 2$, $\mathcal{N}_2 = 4$, $\mathcal{N}_3 = 8$ and $\mathcal{N}_4 = 15$. This allows as well for a derivation of the exponential growth

of the number of sectors $\mathcal{N} \sim b^L$ with

$$b = \frac{1 + \sqrt{3 + 2\sqrt{5}}}{2} \approx 1.87.$$

The size of the largest sector does not have a readily apparent combinatorial description but can be identified as the sequence numbered A073028 in Ref. 20. It has the approximate form

$$N \sim \frac{5^{1/4}\,\varphi^L}{\sqrt{2\pi(L+1)}}, \text{ where } \varphi = \frac{1+\sqrt{5}}{2}.$$

A similar numerical investigation into the sectors of the $U(1)$ East transfer matrix reveals that each sector has a unique representative where no particle can hop to the right and thus avoids configurations of the form ●∘●. This leads to a recursion relation for the number of sectors $\mathcal{N}_L = \mathcal{N}_{L-1} + \mathcal{N}_{L-2} + 1$, with $\mathcal{N}_1 = 2$, $\mathcal{N}_2 = 4$. This results in a Fibonacci form for the number of sectors $\mathcal{N}_L = F_{L+2} - 1$ (where $F_n$ is the $n$th Fibonacci number), and an exponential growth, $\mathcal{N}_L \sim \varphi^L$, with $\varphi$ being the golden ratio.

The sequence numbered A101461 in Ref. 20 corresponds to the size of the largest sector; its exact, explicit binomial representation is

$$N = \frac{m+1}{L+1}\binom{L+1}{\frac{L-m}{2}} \text{ where } m = \left\lfloor \sqrt{L+2} - \frac{1+(-1)^{\lfloor L+\sqrt{L+2}-1 \rfloor}}{2} \right\rfloor .$$

The growth of this function takes the asymptotic form $N \sim 2^L/L$. The charge of the states in the largest sector is $(L-m)/2$, and thus, the largest sector consists of a fraction $(m+1)/(L+1) \sim 1/\sqrt{L}$ of states with the same charge.

### 5.2. XNOR constraint

The "XNOR" constraint allows hopping between sites $j$ and $j+1$ if sites $j-1$ and $j+2$ are *both* occupied or both unoccupied. Similar constraints have been considered in Refs. 21–26. Like the fragmented models discussed above, the XNOR model has an exponential number of sectors that are not coupled by the dynamics; unlike PXYP and $U(1)$ East, the XNOR model additionally has a large number of sectors in which the dynamics can be solved exactly, giving a gap $\Delta \sim L^{-z}$, with $z = 4$. These sectors correspond to a single magnon-like excitation hopping randomly in an otherwise-frozen background configuration. While the magnon's motion is diffusive (i.e., a random walker), it transports no net charge on average, as its charge flips sign as it traverses domains of alternating charge (the magnon is particle-like in a frozen hole domain, and hole-like in a frozen particle domain).

We argue that this "screening" of the magnon's charge gives a twofold enhancement of $z$ for charge correlations as compared to the underlying magnon's position (in this case diffusive, with $z = 2$); this prediction of $z = 4$ from screening matches the scaling of the gap. While our exact solutions only apply to sectors with a single magnon-like excitation, typical sectors can be described as having many magnons with the same properties; thus, we predict that the infinite-temperature correlation function, $\mathcal{C}(x,t)$, computed in the main text should reflect $z = 4$ subdiffusion.

*a. Symmetries and decoupled sectors.—* A special property of the XNOR model is that, in addition to charge, the number of domain walls is also conserved by the allowed moves (here, a domain wall exists between sites $j$ and $j+1$ if their occupations differ). Additionally, the allowed moves constrain the motion of the domain walls themselves, so that isolated domain walls are frozen, but adjacent pairs of domain walls (corresponding to isolated magnons in some background) can hop. Thus, there exist an exponential number of completely frozen states that have no adjacent domain walls. In these frozen states the configuration is such that particles and holes only appear in adjacent pairs or larger contiguous clusters, and never in isolation.

As for the fragmented PXYP and $U(1)$ East models, brute force enumeration of states on system sizes up to $L = 20$ allows for the identification of sequences describing both the number of sectors and the maximum

sector size. The maximum sector size for a chain of $L$ sites is exactly the same as for PXYP, and corresponds to sequence A073028 in Ref. 20. As previously noted, this sequence grows exponentially with a base of the golden ratio, and thus represents an exponentially decaying fraction of the total Hilbert space size. The number of sectors is $\mathcal{N}_L = 2\,F_{L+1} - 2$, and thus also grows exponentially as $\varphi^L$. This counting of sectors suggests that the model is strongly fragmented, and by the criteria considered in the previous section we would expect correlators to localize. Numerics on the correlator (shown in the main text) shows that it does not in fact localize, but instead exhibits $z = 4$ subdiffusion.

  b.  *Exact solution of single magnon sectors.—*  To explore why subdiffusion with $z = 4$ occurs in this model, we will exactly solve the dynamics in some sectors. As discussed above, configurations in which particles and holes always appear in strings of length two or more are completely frozen, and thus the these sectors are dynamically trivial. The solvable sectors consist of states that differ from one of these frozen configurations by a single spin flip. For example, the state

$$|\circ\circ\circ\circ\circ\circ\bullet\bullet\bullet\bullet\bullet\circ\circ\circ\bullet\bullet\bullet\bullet\rangle \ ,$$

is frozen while the state

$$|\circ\circ\bullet\circ\circ\circ\bullet\bullet\bullet\bullet\bullet\circ\circ\circ\bullet\bullet\bullet\bullet\rangle$$

has a single mobile particle in the leftmost domain. This mobile particle (magnon) can hop to the left or right until it abuts a domain of particles. For example, after two hops of this particle to the right, the resulting configuration is

$$|\circ\circ\circ\circ\bullet\circ\bullet\bullet\bullet\bullet\circ\circ\circ\bullet\bullet\bullet\bullet\rangle \ .$$

At this point, the particle can no longer hop to the right, but the hole on the site immediately to its right *can* hop, resulting in

$$|\circ\circ\circ\circ\bullet\bullet\circ\bullet\bullet\bullet\bullet\circ\circ\circ\bullet\bullet\bullet\bullet\rangle \ .$$

Now, this hole is the only degree of freedom that can move: In a sense, the magnon (i.e., mobile excitation) has changed sign (from a particle to a hole). This hole-like excitation can hop until it hits the right boundary of the domain, at which the configuration again resembles a free particle-like mover to the right of a frozen particle domain,

$$|\circ\circ\circ\circ\bullet\bullet\bullet\bullet\bullet\circ\bullet\circ\circ\circ\bullet\bullet\bullet\bullet\rangle \ .$$

This entire sequence of moves can be viewed as a single, unimpeded, magnon-like excitation travelling as a particle through empty domains and as a hole through filled domains. Each configuration in such single-magnon sectors is dynamically connected to exactly two other states, in which the magnon has hopped to the left/right of its current position. As argued above, the magnon interacts with the otherwise-frozen background domains: As the magnon first hops through a domain from left to right, the position of the domain itself shifts two sites to the left, which has a marked effect on the energy spectrum of the magnons.

  To see this, we construct states $|m,\mathbf{b}\rangle$, where $m$ labels the position of the magnon and $\mathbf{b} = (b_1, b_2, ...b_n)$ labels the positions of the leftmost site of each domain (both particle- and hole-like), and we identify the operator $\tau$ that hops the magnon to the right,

$$\tau|m,(b_1,b_2,\ldots,b_n)\rangle = \begin{cases} |m+2,(b_1,\ldots,b_i-2,\ldots,b_n)\rangle & m = b_i - 3 \\ |m+1,(b_1,\ldots,b_n)\rangle & \text{otherwise.} \end{cases} \quad (5.1)$$

With periodic boundary conditions on a chain of $L$ sites, we find that

$$\tau^{L-n-2}|m,(b_1,\ldots,b_n)\rangle = |m-2,(b_1-2,\ldots,b_n-2)\rangle = T^{-2}|m,(b_1,\ldots,b_n)\rangle \ ,$$

where $T$ is the translation operator that shifts everything one site to the right. Thus, in the single-magnon sector with $n$ domain walls, $\tau^{L-n-2} = T^{-2}$. As the number of domain walls, $n$, is always even, we find for even $L$ that

$$\tau^{\frac{1}{2}(L-n-2)L} = T^{-L} = \mathbb{1} \ ,$$

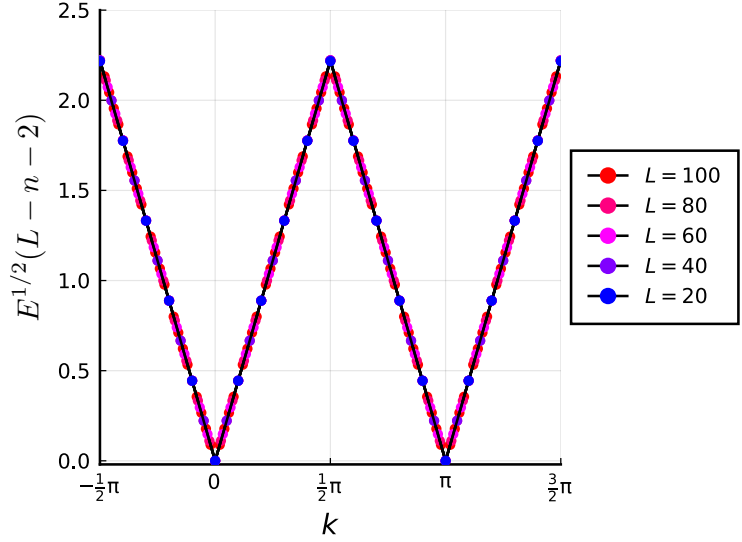

FIG. 7. **Spin wave spectrum of the effective XNOR Hamiltonian.** Using exact diagonalization (ED), we plot the rescaled magnon dispersion relation (5.2), in the sector with $n = 4$ domain walls, finding agreement with ED data.

giving $N = L(L-n-2)/2$ states in this single magnon sector; the magnon has to hop around the boundary $L/2$ times in order for the initial configuration to come back to itself.

The dynamics within each single magnon sector can be described as an effective single particle hopping problem on $N$ sites. The RK Hamiltonian restricted to this sector takes the form $H = 1 - \frac{1}{2}(\tau + \tau^\dagger)$. The energy eigenstates are

$$|\kappa\rangle = \sum_{y=0}^{N-1} e^{i\kappa y} \tau^y |m, \mathbf{b}\rangle$$

and the associated energy eigenvalues are $E(\kappa) = 1 - \cos(\kappa)$, with $\kappa$ quantized as $0, 2\pi/N, 4\pi/N, \ldots, 2\pi(N-1)/N$. By computing the action of $T$ on these states, we see that the physical momentum, $k$, is related to the pseudomomentum, $\kappa$, according to

$$k \equiv \frac{L-n-2}{2}\kappa \quad \mod 2\pi.$$

The lowest energy band of states has

$$\kappa = \frac{2}{L-n-2}k \;,\; \text{with} \;\; E(k) = 1 - \cos\frac{2k}{L-n-2}\;.$$

For large $L$ (or small $k$), the dispersion is

$$E(k) \approx \frac{1}{2}\left(\frac{2k}{L-n-2}\right)^2, \tag{5.2}$$

see Fig 7. Note that the lowest excited state has $k = 2\pi/L$ and an energy gap $\Delta \sim L^{-z}$ with $z = 4$. In summary, the quantization condition for the single magnon states leads to a gap with subdiffusive scaling, despite the fact that the magnons themselves execute a random walk, corresponding to ordinary diffusion.

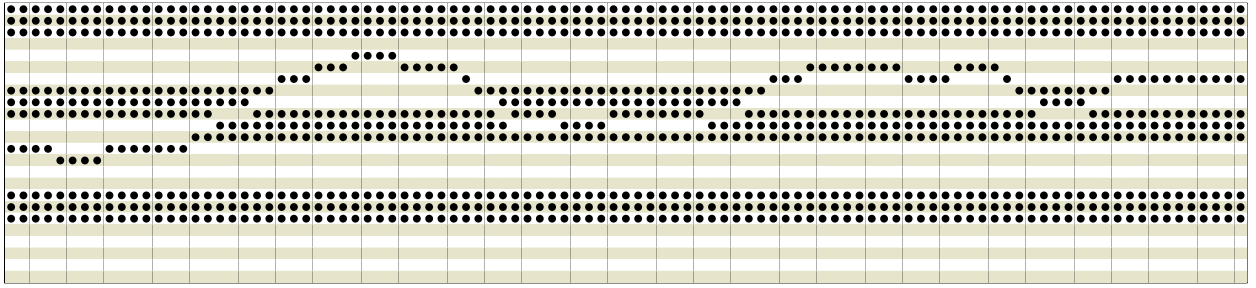

FIG. 8. **Diffusive magnons in the XNOR model.** Trajectory of a single magnon (seen as a lone black dot in the leftmost "column") moving in a background of frozen domains. The horizontal axis is time and the vertical axis is position. As the magnon traverses the middle particle-like domain, its charge changes (it becomes hole-like) and the frozen domain's position shifts by two sites opposite the magnon's propagation direction.

*c. Screening argument.—* The origin of the subdiffusive dynamical exponent, $z = 4$, can also be justified intuitively by generalizing the "screening" argument of Ref. 27. This argument provides an explanation for the finite temperature diffusive spin transport in the integrable XXZ chain in the easy-axis regime, at half-filling (no net magnetic field). As the XXZ spin chain is integrable, it supports stable magnon-like quasiparticles at arbitrary temperature, which move ballistically. One would then expect spin transport to be ballistic as well, but the charge of the magnons happens to be screened dynamically as they move through the half-filling environment, leading to diffusion [27–29]. In our case, we have *diffusive* magnons: in a single magnon sector considered above, the magnon excitation is simply performing a random walk. However, its charge changes as it moves through the system: the magnon is either a *particle* as it moves through the vacuum, or a *hole* as it moves through a frozen domain (see Fig. 8). Therefore, the magnon transports no net charge on average, as its charge flips sign as it traverses domains of alternating charge. Generalizing the arguments of Ref. 27, such diffusing magnons with fluctuating charge precisely lead to subdiffusion with $z = 4$ [30]. The consequences for near integrable quantum spin chains will be reported in Ref. 30.

### 5.3. Fredkin and Motzkin constraints

Here we discuss some properties of the classical dynamics generated by the Fredkin model transfer matrix Eqs. 3.42-3.43. Some of these properties were discussed previously in connection to studies of the Fredkin and Motzkin quantum spin chains [31–34]. As discussed in Sec. 2.3, the "low energy" properties of these transfer matrices correspond to a quantum spin chain Hamiltonian tuned to an unfrustrated RK point. We can thus draw from studies of the Fredkin Hamiltonian to learn about the transfer matrix and its dynamics.

*a. Symmetries and decoupled sectors.—* The Fredkin model has $U(1)$ particle number conservation; the model's constraint explicitly breaks particle-hole symmetry, $\mathcal{C}$, and parity (spatial reflection), $\mathcal{P}$, but conserves the combination, $\mathcal{CP}$. The $U(1)$ symmetry decouples the dynamics between different charge sectors; additionally, the $\mathcal{CP}$ symmetry splits the half-filled charge sector into $\mathcal{CP}$-even and $\mathcal{CP}$-odd states.

With open boundary conditions, each charge sector further splits into sectors that do not couple under the dynamics. To understand these sectors, it is convenient to introduce the following pictorial representation of the Hilbert space [31]: Each basis state (in the occupation basis) is represented by a path, where a hole at site $j$ is depicted as an upward-slanting line segment, $\circ \to \diagup$, and a particle at site $j$ is depicted as a downward-slanting line segment, $\bullet \to \diagdown$. The height of the path at site $j$ encodes the excess of holes versus particles on all sites to the left of $j$. In this representation, the allowed dynamical moves always preserve the minimum height of the path in the vicinity of the move. The path's global minimum height, $m$, is conserved by all allowed moves; states with different $m$ belong to decoupled sectors. For each charge sector, there are $O(L)$ possible values for $m$, resulting in $\mathcal{N} \sim O(L^2)$ decoupled sectors. More precisely, $\mathcal{N} = (L^2 + 4L + 4)/4$ sectors for even $L$ and $\mathcal{N} = (L^2 + 4L + 3)/4$ for odd $L$.

With periodic boundary conditions, the minimum path height, $m$, is no longer conserved: The heights of the path shift everywhere when a particle hops around the boundary. For the exactly half-filled sector, a

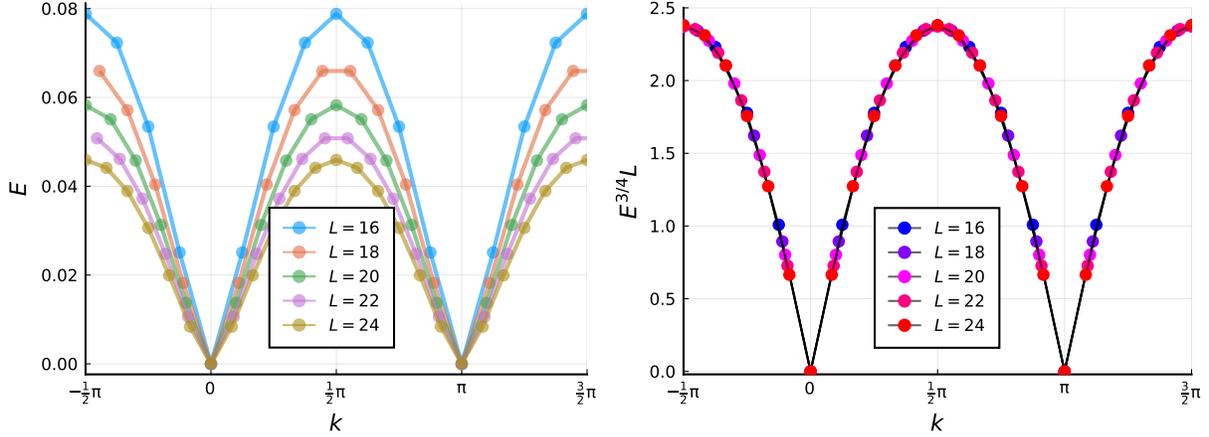

FIG. 9. **Spectrum of the Fredkin Hamiltonian.** *Left*: Lowest lying energy eigenvalues for various momenta, $k$, in the half-filled charge sector, computed using exact diagonalization (ED) for several system sizes. *Right*: The same data rescaled to show $E^{3/4}L$ for each eigenvalue, $E$. The curves collapse neatly, and the data for small $k$ are consistent with $E^{3/4}L \propto |k|$.

particle hopping around the boundary necessarily changes $m$ by $\pm 2$. Thus, the half-filled sector splits into two decoupled sectors, corresponding to even/odd $m$, while in all other charge sectors, all states of a given charge are dynamically connected [31].

  b.  *Low energy spectrum of the Fredkin Hamiltonian.—* Previous studies of the Fredkin Hamiltonian using DMRG on an open boundary condition chain suggest a gap in the zero-charge sector that scales as $E \sim L^{-z}$ with $z \approx 2.69$ [33]. That result used simulations on open chains with $L \sim 200$ sites to extrapolate to infinite size, in an attempt to account for finite-size corrections. To complement that numerical evidence, we study the spectrum of the Fredkin Hamiltonian with periodic boundaries using exact diagonalization on small systems, where momentum and charge quantum numbers can be resolved. The lowest energy eigenvalues in each momentum sector among half-filled states are shown in Fig. 9.

A few features merit discussion: First, there are two $E = 0$ ground states, corresponding to momenta $k = 0$ and $k = \pi$. This twofold degeneracy corresponds to the decoupled even- and odd-$m$ sectors discussed previously. The $k = 0$ ground state is the uniform superposition of basis states with both even and odd $m$, while the $k = \pi$ ground state is a signed superposition thereof, with signs $(-1)^m$ [31]. Second, the energy band has $k \to -k$ degeneracy, a feature guaranteed by the model's $\mathcal{CP}$ symmetry. Finally, we see an unusual scaling of these energies with system size and momentum, suggestive of the form

$$E(k,L) = \left(\frac{f(k)}{L}\right)^{4/3}, \tag{5.3}$$

with $f(k) \propto |k|$ for small $k$. As the smallest allowed $k$ for system size $L$ is $k_{\min} = 2\pi/L$, this suggests an energy gap $\Delta \propto k_{\min}^{4/3} L^{-4/3} \sim L^{-8/3}$. This is scaling is consistent with the conclusion that the gap scales as $\Delta \sim L^{-z}$ with $z = 8/3$, but in stark contrast to the expectation that $E(k,L) \sim |k|^z$ for the *same z*. The system sizes accessible to ED are fairly small, but this can be remedied in future work by using DMRG with periodic boundary conditions.

The similarities between the Fredkin spectrum (5.3) and that of the XNOR model suggest a possible explanation for the $z = 8/3$ subdiffusion in terms of *screening*. By analogy with XNOR, we expect magnon-like excitations with a superdiffusive dispersion $E(k) \sim k^{z_m}$, where $z_m = 4/3$ for the bare magnons, which transport zero net charge on average, due to the same screening argument as in Sec. 5.2. That argument relates the dynamical exponent of the charges to that of the magnon's motion as $z = 2z_m$; correspondingly the gap should scale as $\Delta \sim L^{-z}$ with $z = 2z_m$ corresponding to *charge* transport. While we do not have an explanation for such a superdiffusive magnon spectrum—or even a description of the magnons as in the XNOR case—this screening hypothesis predicts diffusion away from half filling, because the charge is only

22

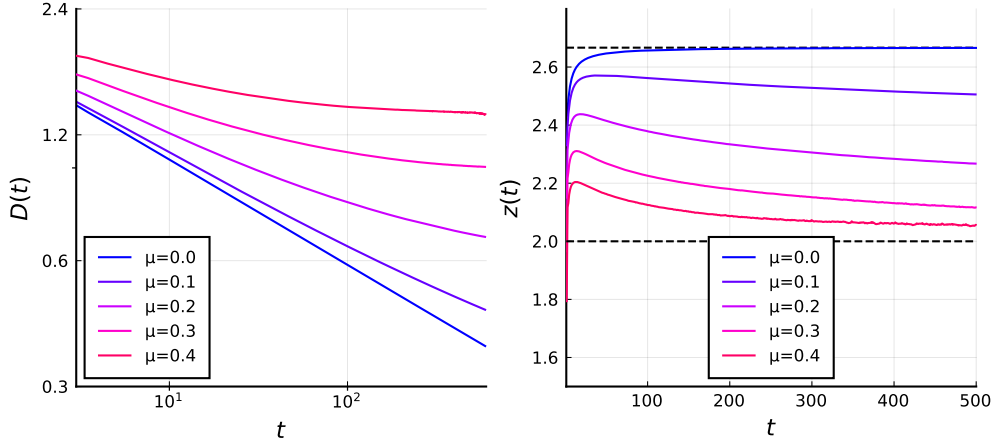

FIG. 10. **Fredkin dynamics in a chemical potential.** *Left*: Effective diffusion constant, $D(t) = \frac{1}{2}dV(t)/dt$, where $V(t)$ is the variance of the charge correlator. The data suggest that the diffusion constant saturates for nonzero $\mu$, while it scales as a power law in $t$ for $\mu = 0$, as expected for subdiffusive transport. *Right*: Effective dynamical exponent extracted from the variance of the charge correlator as a function of chemical potential, $\mu$. The curves suggest a convergence to $z = 2$ for *any* nonzero $\mu$.

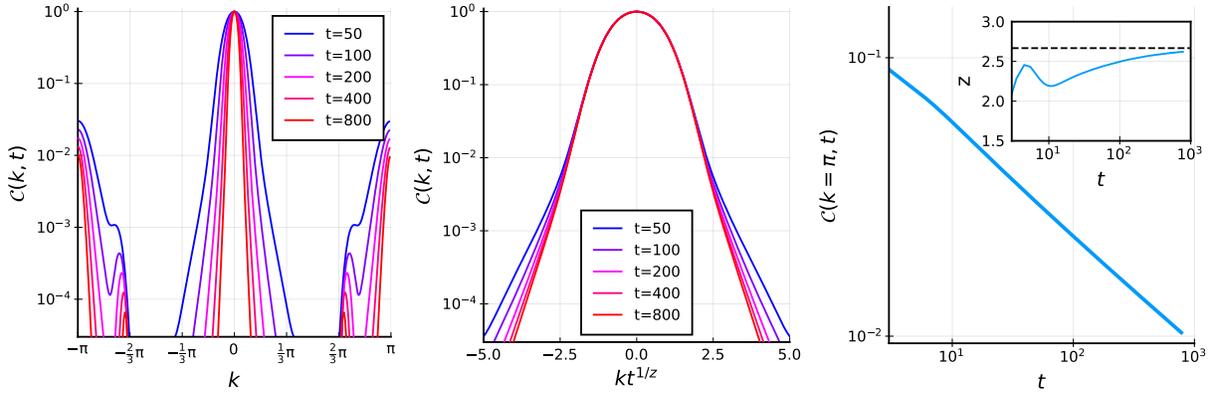

FIG. 11. **Momentum-space structure factor, $\mathcal{C}(k,t)$, for the Fredkin model.** *Left*: Fourier transformed two-point correlation function computed from $\mathcal{C}(x,t)$ as plotted in the main text. The data show a stable peak at $k = 0$ whose width narrows over time as $t^{-1/z}$, and a secondary decaying peak at $k = \pi$ whose height decays as $t^{-1/z}$. *Middle*: The same data plotted with a rescaled width, $kt^{1/z}$ with $z = 8/3$, to show a collapse near $k = 0$. The collapse function takes the approximate form $\mathcal{C}(k,t) \sim \exp(-|k|^z t)$ for small $k$ (see main text for another view) but deviates for larger $k$. *Right*: Height of the $k = \pi$ peak shown on a log scale, showing a power-law decay. *Inset*: The exponent $z$ governing the power-law decay, $\mathcal{C}(k = \pi, t) \sim t^{-1/z}$, approaches 8/3.

partially screened away from half filling, a prediction that can be tested numerically. The results are shown in Fig. 10: The data suggest that any nonzero chemical potential leads to diffusive transport, consistent with the screening hypothesis.

  *c. Momentum space view of the Fredkin structure factor.—* In Fig. 11, we show more details of the Fredkin correlation function $\mathcal{C}(x,t)$ by Fourier transforming the data, producing the momentum-space structure factor $\mathcal{C}(k,t)$. As can be seen in the figure, the structure factor consists of two main peaks at $k = 0$ and $k = \pi$. Both peaks become sharper over time, with widths that scales like $t^{-1/z}$. However, the $k = \pi$ peak slowly decays away, with a height that scales like $t^{-1/z}$. Thus, the asymptotic scaling function $f(u)$, where $\mathcal{C}(x,t) \sim 1/t^{1/z} f(x/t^{1/z})$, should ultimately only have contributions from the $k = 0$ peak.

23

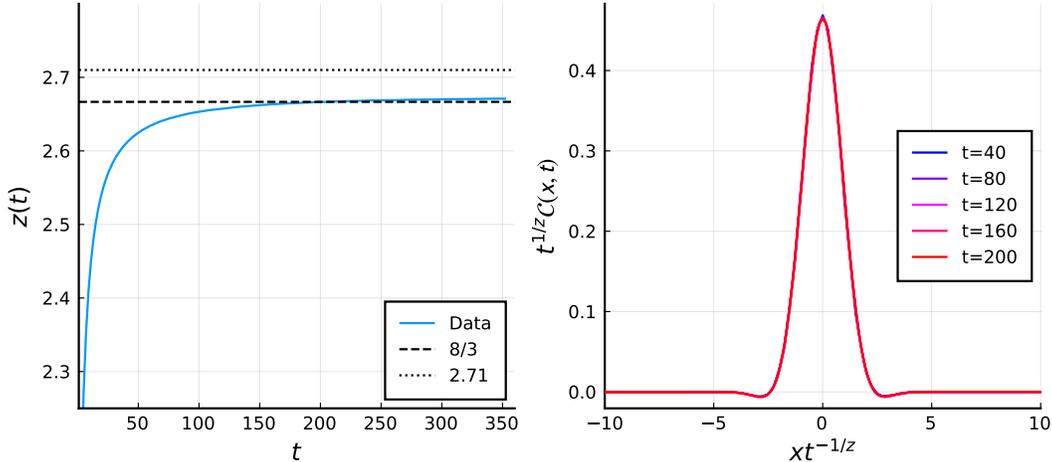

FIG. 12. **Motzkin random circuit.** Properties of the Haar averaged correlation function $\mathcal{C}(x,t) = \mathcal{D}^{-1}\,\mathrm{tr}\left[\overline{S_0^z(t)\,S_x^z(0)}\right]$ in the Motzkin circuit, as computed via TEBD of the transfer matrix. *Left*: The dynamical exponent, $z(t)$, as a function of time; the exponent appears closer to (albeit slightly above) our prediction of $8/3$ (dashed line) than the value $2.71$ (dotted line) reported in Ref. 33. *Right*: The correlation function collapsed to the scaling form $\mathcal{C}(x,t) \sim 1/t^{1/z} f(x/t^{1/z})$ with $z = 8/3$.

The collapsed $k = 0$ peak approximately takes the shape $\exp(-|k|^z t)$; however, there are significant deviations for larger $k$, which may or may not disappear at later times. This may provide a hint for further studies aiming to identify the hydrodynamic equations governing Fredkin-like dynamics or otherwise derive the subdiffusive behavior from first principles.

*d. Numerical evidence for the universality of $z = 8/3$ subdiffusion.—* To demonstrate the universality of $z = 8/3$ subdiffusion, we first consider a model closely related to Fredkin, known as the "Motzkin" model. Like Fredkin, the Motzkin constraint can be expressed in terms of a path representation where the allowed moves do not change the local minimum of the height field. The precise form of the constraint is detailed in Sec. 3.8.

The associated Hamiltonian for this model has been studied in the low-temperature setting alongside the Fredkin model in Ref. 33, finding $z \approx 2.71$; Motzkin dynamics were considered in Ref. 36, which found $z \approx 2.5$. Repeating the calculations shown in the main text for the Fredkin model yields the results shown in Fig. 12: The variance scales as $\sigma^2(t) \sim t^{2/z}$ with $z$ approaching $8/3$, and the Haar averaged correlation function, $\mathcal{C}(x,t) = \mathcal{D}^{-1}\,\mathrm{tr}\,[S_0^z(t)S_x^z]$, collapses nicely to the scaling form $\mathcal{C}(x,t) \sim 1/t^{1/z} f(x/t^{1/z})$ with $z = 8/3$. The asymptotic value of $z(t)$ is close to $8/3$ ($z(t = 350) \approx 2.67$). Compared to estimates of $z$ from DMRG calculations, the TEBD calculation lacks finite-size corrections due to boundary effects and allows for the use of the full scaling collapse to improve estimates of $z$. In addition to the computation from the variance of the charge profile, the dynamical exponent $z$ can also be computed from the return probability $\mathcal{C}(x=0,t) \sim t^{-1/z}$, as shown in Fig. 5 for the Fredkin model, with comparable precision.

Next, we consider another Fredkin-like circuit model constructed by enlarging the size of the circuit gates to four sites and allowing Haar random mixing between any states with the same minimum height in the path representation. Enlarging the gates has been previously shown to affect the dynamics in fracton circuits exhibiting fragmentation for smaller gate sizes [17]. The expanded gate model also exhibits $z \simeq 8/3$ subdiffusion as can be seen in the profile collapse and the dynamical exponent's behavior as a function of time, $z(t)$, in Fig. 13, and exhibits the same scaling form as the Fredkin and Motzkin circuits.

As a final check, we considered the Fredkin dynamics under purely stochastic classical simulations. Each simulation started from a uniformly randomly chosen configuration of occupied and empty sites, and at each circuit gate a jump to another basis state occured with probability $1/2$ if the constraints allowed. In this formulation, each random instance of the dynamics is extremely cheap to compute — however the correlation

24

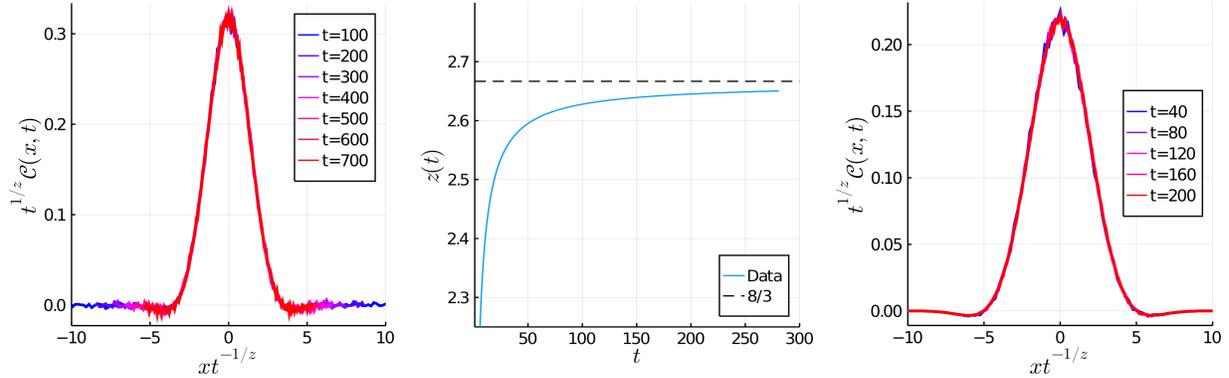

FIG. 13. **Universality of $z = 8/3$ for Fredkin-like models.** The scaling form for correlators under Fredkin dynamics appears as well for Fredkin variants, suggesting that this exponent is universal. *Left*: Profile collapse of stochastic Fredkin simulations; the correlation function is averaged over $2 \times 10^7$ random instances. *Middle*: The dynamical exponent, $z(t)$, as a function of time for Fredkin-like dynamics with four-site gates; the exponent reaches $z \approx 2.65$ out to the latest times simulated, but appears to continue increasing to the predicted value of $8/3$ (dashed line). *Right*: Profile collapse of correlators for Fredkin-like dynamics with four-site gates and $z = 8/3$.

function must be averaged over tens of millions of instances to capture the scaling profile as well as in TEBD.

25



\end{document}